\documentclass[%
 aip,
 amsmath,amssymb,
 reprint,%
]{revtex4-1}

\usepackage{graphicx}
\usepackage{dcolumn}
\usepackage{bm}
\usepackage{graphicx}
\usepackage{dcolumn}
\usepackage{bm}
\usepackage{multirow}
\usepackage{csquotes}
\usepackage{xparse}
\usepackage{tabu} 
\usepackage{xparse,xcoffins}
\usepackage{dcolumn}
\usepackage{bm}
\usepackage{caption}
\usepackage{booktabs}
\usepackage{float}
\usepackage{color}
\usepackage{soul}
\usepackage{xcolor}
\usepackage{subcaption}
\usepackage{chemformula} 
\usepackage[T1]{fontenc} 

\makeatother
\begin{document}

\preprint{AIP/123-QED}
\title{\centering Effects of a Flexible Ion Gel as an Active Outer-Layer when in Contact with a Metallic Electrode}
\author{Elton A. de Moura}%
\affiliation{Universidade Federal do Paran\'a - UFPR, Curitiba, Paraná, Brazil}
 \author{Ana C. de Paula}
 \affiliation{Universidade Federal do Paran\'a - UFPR, Curitiba, Paraná, Brazil}
\author{Adriano R. V. Benvenho}
 \email{adriano.benvenho@ufabc.edu.br}
\affiliation{Universidade Federal do ABC - UFABC, Santo Andr\'e, S\~ao Paulo, Brazil}
\author{José P. M. Serbena}
  \email{serbena@fisica.ufpr.br}
\affiliation{Universidade Federal do Paran\'a - UFPR, Curitiba, Paraná, Brazil}
\author{Keli F. Seidel}
  \email{keliseidel@utfpr.edu.br}
\affiliation{%
Universidade Tecnol\'ogica Federal do Paran\'a - UTFPR, Curitiba, Paran\'a, Brazil
}%

\date{\today}

%

                       %
\begin{abstract}
In this work the effect of an ion gel outer-layer stuck on top of ITO/PBT/Sn devices was investigated towards its effects on the electrical properties. 
When 
this external electrolyte film is in contact with any top permeable electrode 
it produces a self-biasing effect and changes the charge carriers injection properties. 
The outer-layer promoted situations where the output current increases up to two orders of magnitude and others where the output current decreases one order of magnitude in comparison to the same samples without it. Admittance spectroscopy measurements were made and the proposed equivalent circuit model indicates that the interfacial electrical properties dominate charge injection face to the bulk properties when the outer-layer is present. All the changes observed here are reversible after the ion gel is detached and replaced, indicating that ions do not diffuse into the active layer. The observed results can contribute to improve the current density in certain sandwich structures as well as notify that electrolyte external films can behave as an active layer promoting electrical changes into sandwich devices and can be extended to cases where the electrolyte film is used as substrate. 

\end{abstract}
%

\maketitle
%

\section{\label{introduction} Introduction}
\par Many optoelectronic devices are based on vertical stacked layers forming a sandwich structure, such as light-emitting diodes (LEDs) \cite{influence_of_charge_injection}, light-emitting electrochemical cells (LECs) \cite{Light_Emitting_Electrochemical_Cells_Heeger, solid_state_electrolytes_polymer_light_emitting_Emil}, electrochromic devices \cite{electrochromic_device},solar cells \cite{review_solar_cell}, vertical transistors \cite{Seidel2018_ambipolar_gate, Tessler_patterned_electrode_lito_2009}, among others. The efficiency of these devices depend on many different conditions. One is regarding to the control of the density of charge carriers available inside the active layer. Different electrical properties could provide higher or lower charge carrier density flowing throughout the device as, e.g, charge carrier mobility from the semiconductor or energy alignment at the interfaces. This second example can be improved by playing with the energy level matching at the interface of two different materials that can be semiconductor/semiconductor (S/S) or metal/semiconductor (M/S) interfaces \cite{interlayer_2019, interlayer_Norbert_2020, interlayer_Semiconductor_semiconductor, review_interlayer_tandem_structure}.
\par The addition of an interlayer between M/S interfaces is a common technique to improve charge carriers injection\cite{Qiu2003, interlayer_Norbert_2020, interlayer_2019}. It changes the energy barrier for charge carriers injection, creating a metal/interlayer/semiconductor (M/I/S) structure \cite{Cattin2009}. The M/I/S is more frequently obtained from sequential layers deposition (or step-by-step preparation) \cite{interlayer_step_by_step_preparation_and_self_organization_2018}. It is also possible to obtain this structure from blend solution composed by the active layer and interlayer materials, where the interlayer material spontaneously segregates onto the electrode (cathode or anode) forming the M/I/S structure \cite{interlayer_step_by_step_preparation_and_self_organization_2018, Seidel_2020_cathode_interlayer}. The deposition procedure of an interlayer presents difficulties like the compatibility with the previous and/or coming layer regarding to solvent, temperature, deposition techniques and accuracy control of very thin thickness, besides to increase production time and cost of fabrication.
\par Latest studies depict that the interlayer and/or the active layer can be added by ionic species resulting in ionic-electronic doping \cite{ionic_electronic_doping_2018, ref2_Electrolyte_Gated_Polymer_Transistors, ref2_Design_rules_for_LECs, ref2_Electrochemical_tuning_of_vertically_aligned_MoS2}. It promotes a  significant band bending change in the semiconductor electronic structure. However, it can result in ions diffusion throughout the semiconductor that promotes ionic doping and/or electrochemical processes. The control of these situations is still challenging, once it can promote oxidation (or reduction) states at the electrodes/semiconductor interface and, consequently, degradation of the semiconductor \cite{ionic_electronic_doping_2018,review_A_Decade_of_Iontronic_Delivery_Devices}.
\par In this work it is proposed that the electrical properties of a metal/organic semiconductor interface can be modify by ionic species localized not as an interlayer at this interface but as an outer-layer. The outer-layer is composed by a flexible ion gel film and it is stuck on top of the sandwich structure metal/organic semiconductor/metal. The ion gel outer-layer is able to change the charge injection throughout the device increasing or decreasing the output current intensity depending on the thickness of the electrode in contact with the ion gel. This external and flexible electrolyte layer is formed by an ionic liquid immobilized inside a polymer matrix and, as it is an outer layer, the ions diffusion into the semiconductor is avoided. It is expected that the analysed effects here can be observed in any device with an ion gel film stuck on any top electrode (or under any bottom electrode). This phenomenon was analyzed as a self-biasing effect \cite{self_biasing_2017} able to improve (or reduce) the efficiency of charge carriers injection throughout sandwich structure devices.
\section{MATERIALS AND METHODS}
\subsection{Materials and films preparation}
\begin{figure*}[htb]
 \begin{subfigure}{0.28\textwidth}
    \includegraphics[width=\linewidth]{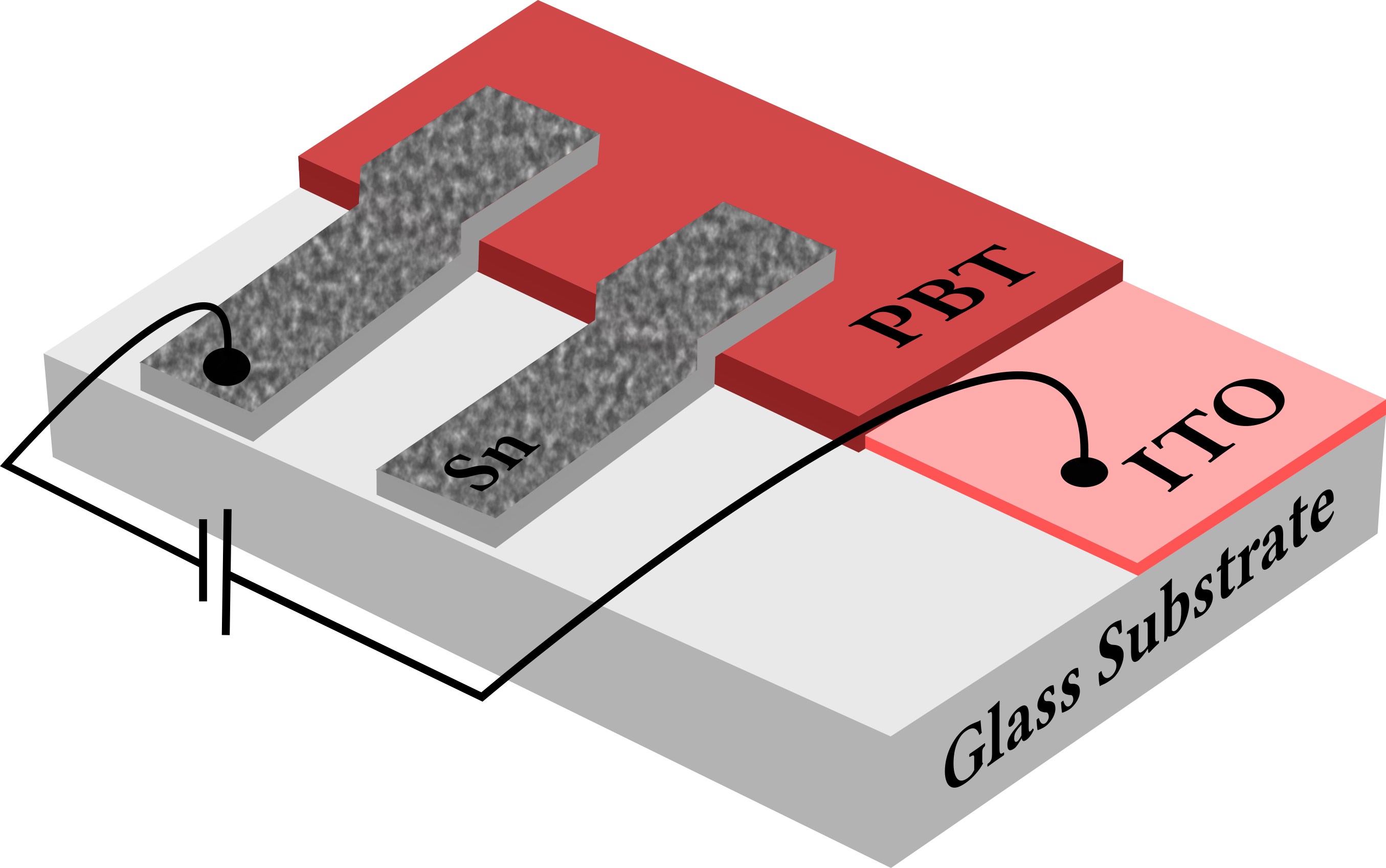}
    \caption{} \label{fig:1a}
  \end{subfigure}%
 \hspace*{\fill}
 \begin{subfigure}{0.28\textwidth}
    \includegraphics[width=\linewidth]{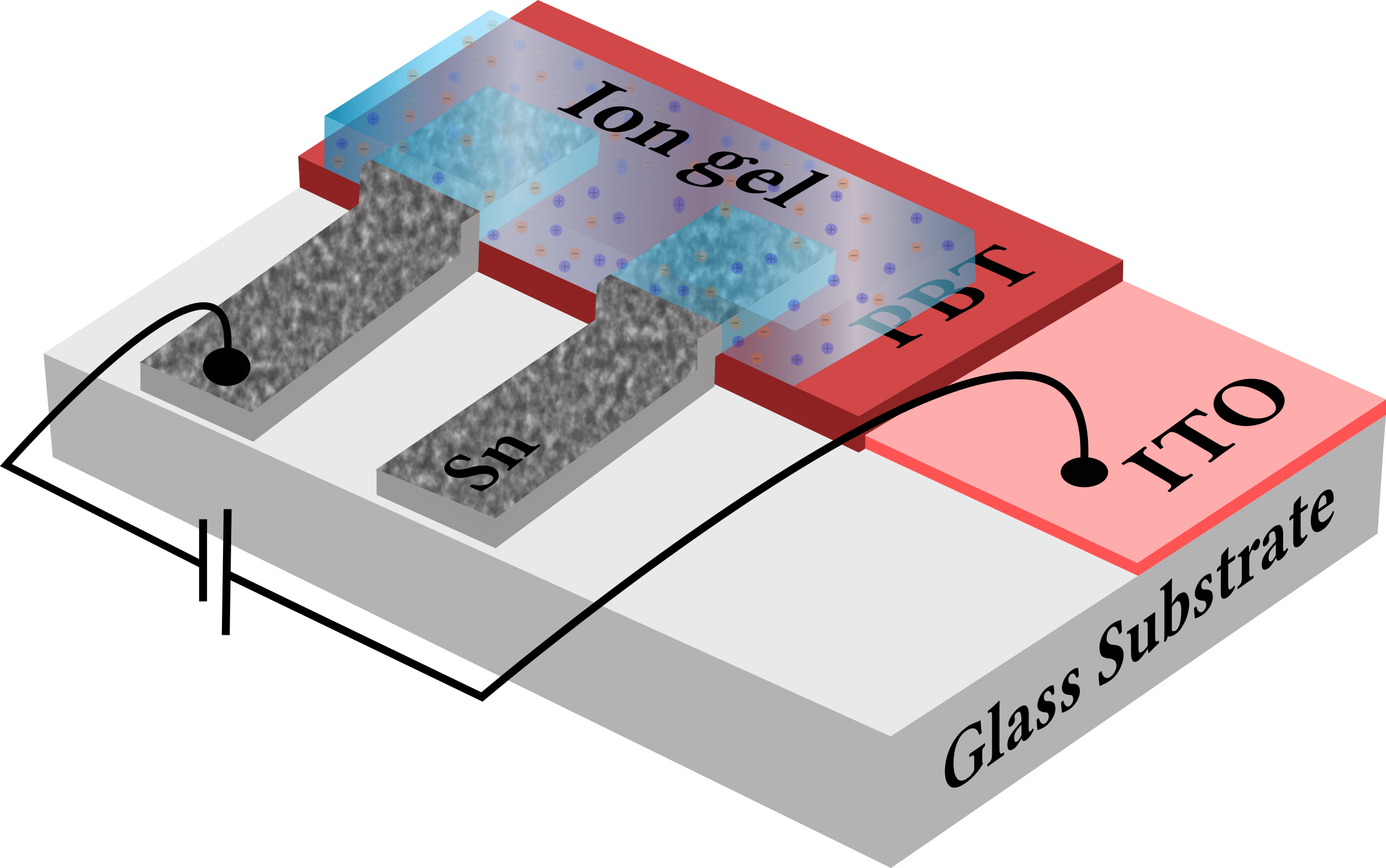}
    \caption{} \label{fig:1b}
  \end{subfigure}%
  \hspace*{\fill}
  \begin{subfigure}{0.28\textwidth}
    \includegraphics[width=\linewidth]{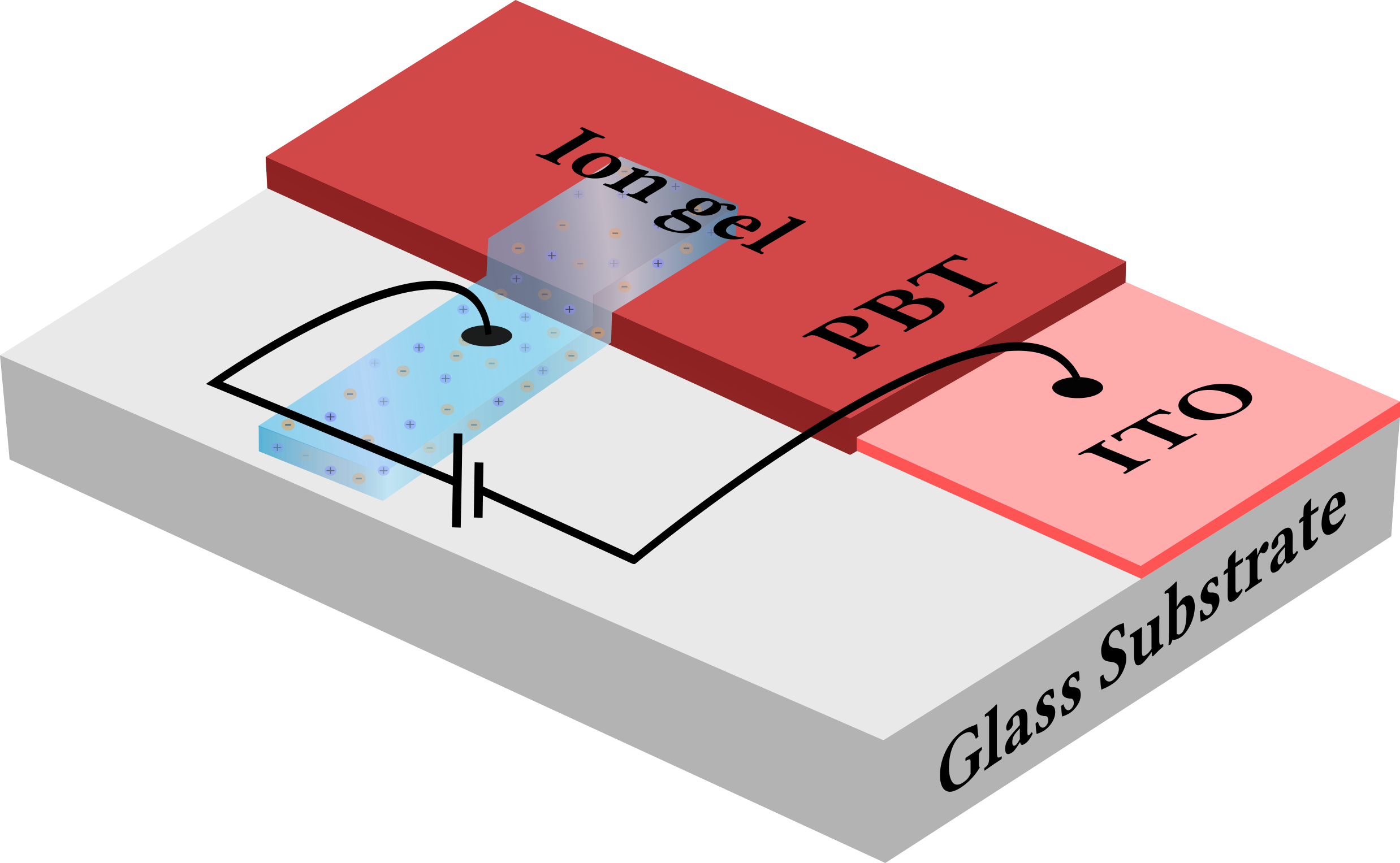}
    \caption{} \label{fig:1c}
  \end{subfigure}%
\caption{Devices structures: (a) ITO/PBT/Sn; (b) ITO/PBT/Sn/ion gel and; (c) ITO/PBT/ion gel, where the ion gel replace the role of the top electrode to provide us a reference-line output current.}
\label{structure/architecture}
\end{figure*}
\par The bottom metal electrode is composed by indium tin oxide (ITO) with 15 $\,\rm{\Omega /square}$ sheet resistance purchased from Luminescence Technology Corp. (LUMTEC). In the sequence, a $250\,\rm{nm}$ thick poly-bithiophene (PBT) is electrochemically synthesized over ITO, based on the same procedure described elsewhere \cite{Souza2014}. The top permeable electrode is composed by an evaporated Sn layer with a deposition rate of $\sim 15\, \rm{s^{-1}}$, controlled during deposition with a crystal quartz oscillator. Three different thickness layers were obtained and studied. 
\par The electrolyte outer-layer is composed by an flexible ion gel film produced from acetone solvent, poly(vinylidene fluoride-co-hexafluoropropylene) (P(VDF-HFP)) and the ionic liquid 1-ethyl-3-methylimidazolium bis(trifluoromethylsulfonyl) amide ([EMIM][TFSI]). The ion gel solution was prepared based on the weight ratio between polymer:ionic liquid:acetone of 1:4:7 \cite{Lee_ion_gel_2012}, respectively. The solution was stirred at $40\, ^{\circ}\rm{C}$ for $40\, \rm{min}$. It was used $130\,\rm{\mu L}$ of ion gel solution drop casted in an area of $\sim (1.3\, \times \, 2)\,\rm{cm^2}$ of a pre-cleaned lime glass slide and awaited 30 min for drying. After that, the ion gel films were cut and stuck on the permeable electrode with tweezers \cite{Lee_ion_gel_2012}, similar and easy as manipulating a sticker. 
\subsection{Device fabrication and characterization}
\par The explored architecture was a vertical sandwich structure metal/polymer/metal composed by ITO/PBT/Sn whose pattern is depicted in Figure \ref{structure/architecture}(a). After that, an ion gel outer-layer was easily stuck on (or removed from) the top of this structure (Figure \ref{structure/architecture}(b)). It was also produced a non-usual device structure with stacked layers of ITO/PBT/ion gel, where the ion gel replaces the role of the top electrode (Figure \ref{structure/architecture}(c)).
\par The electrical characterization was performed with a Keithley 2602 dual-source meter and a Agilent 4284A Precision LCR meter instrument in dark ambient atmosphere and room temperature, setting Sn or ion gel as the common electrode. AC measurements were performed by applying an AC voltage of $0.5\,\rm{V}$ and a null DC voltage. The equivalent electrical circuit simulations of admittance spectroscopy were performed in Matlab considering surfaces/interfaces and bulk properties of the semiconductor as, e.g., traps energy levels. Scanning electron microscopy (SEM) images were obtained in a VEGA3 LMU TESCAN microscopy.
\section{Results}
\par
\begin{figure}[h!]
    \includegraphics[width=0.99\columnwidth]{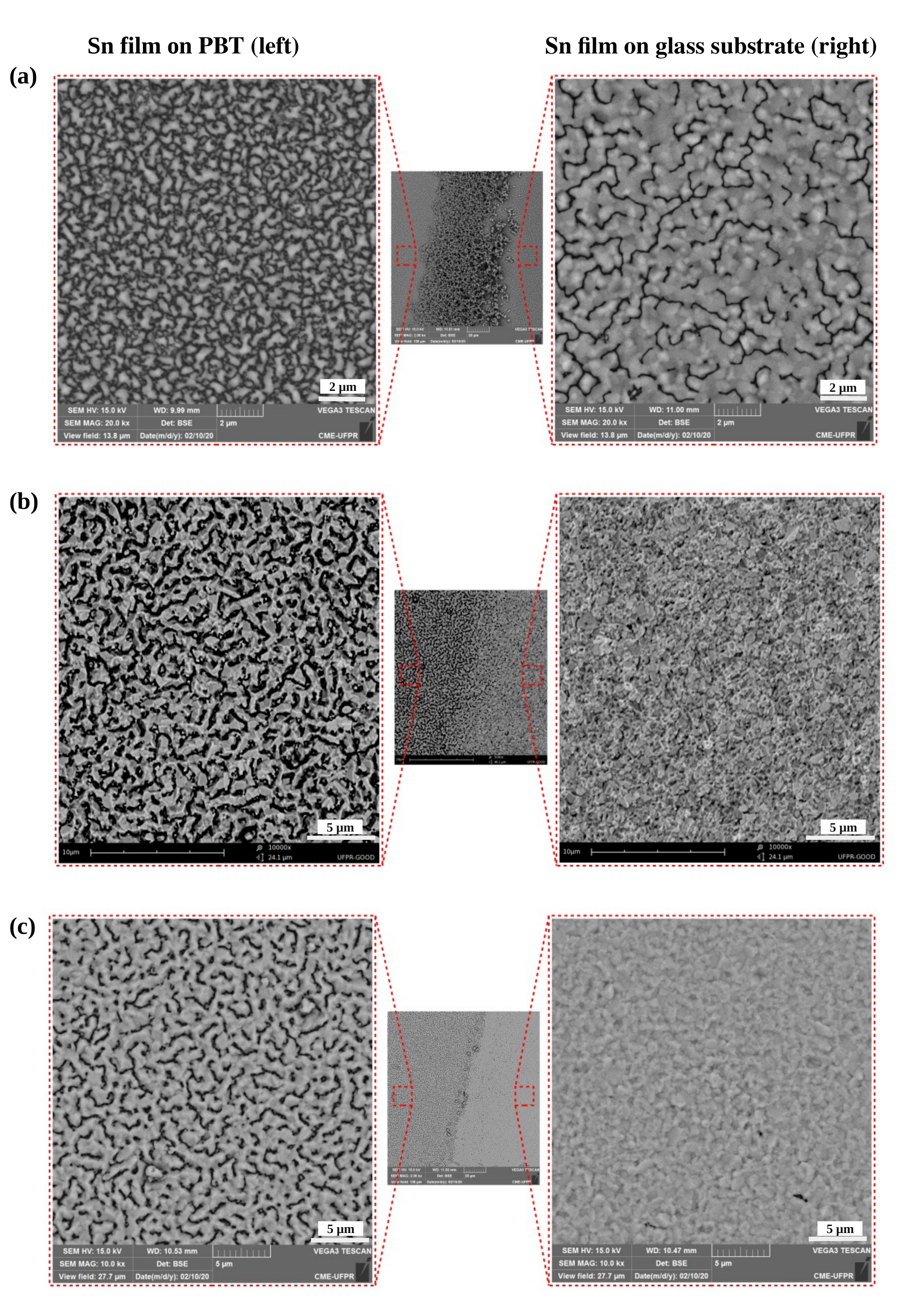}
    \caption{SEM images of evaporated Sn films on top of PBT (left side) and  glass substrate (right side) for different Sn film thicknesses named: (a) thin ($\sim 128\,\rm{nm}$); (b) intermediate ($\sim 264\,\rm{nm}$) and (c) thick ($350\,\rm{nm}$).}
    \label{MEV_images}
\end{figure}

\begin{figure}[h!]
\begin{center}
\includegraphics[width=0.47\columnwidth]{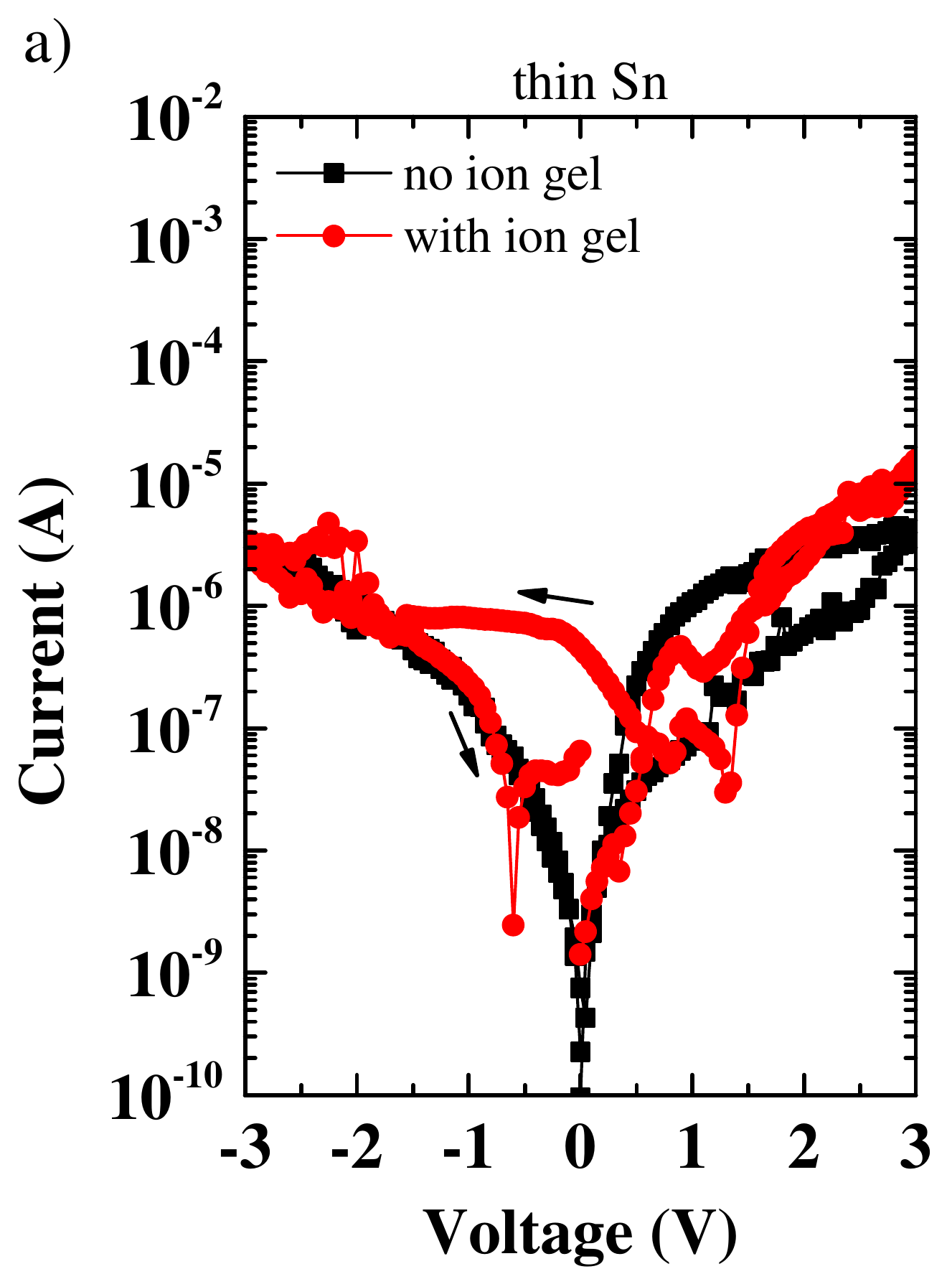}
\includegraphics[width=0.47\columnwidth]{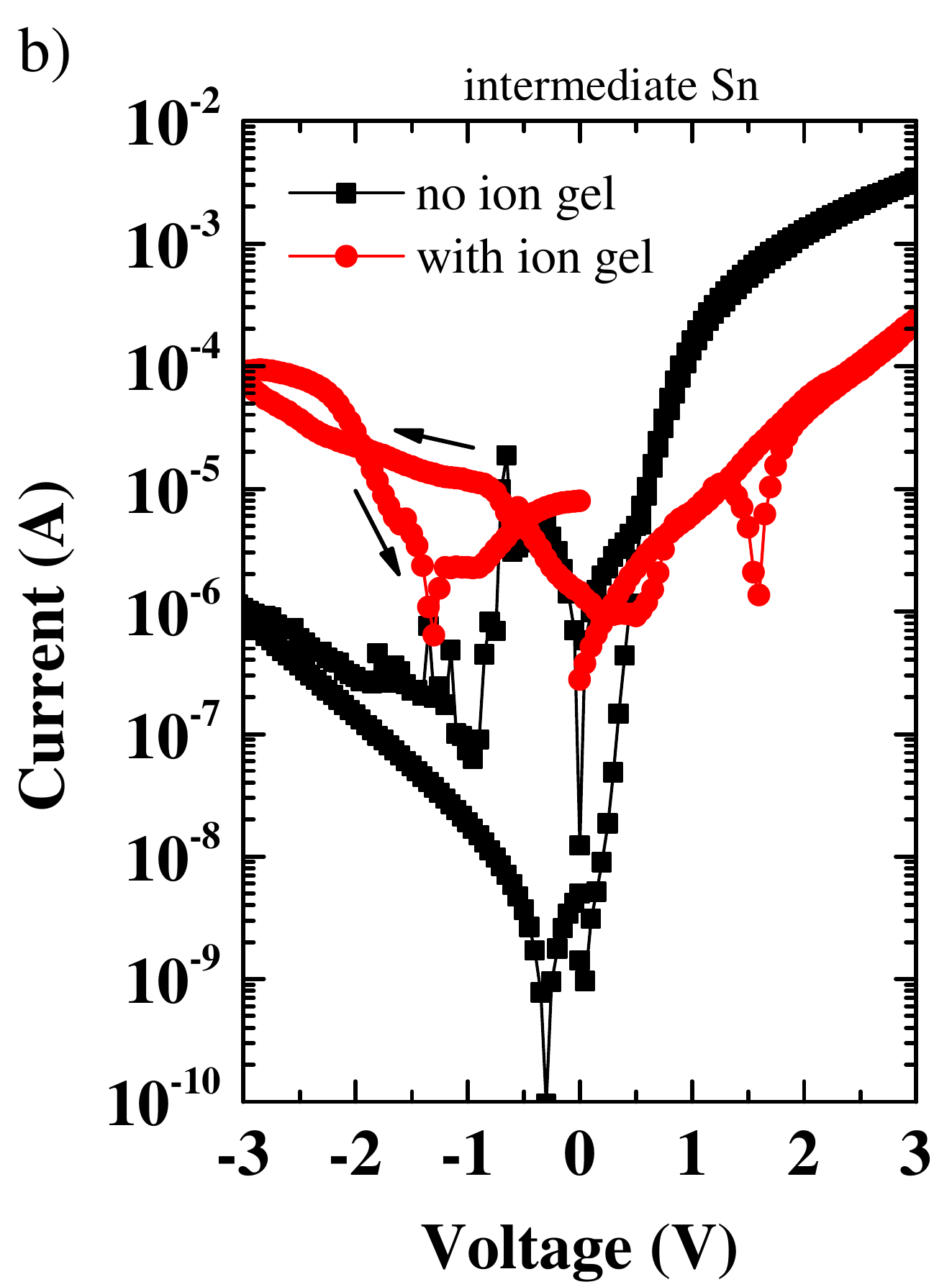}
\includegraphics[width=0.47\columnwidth]{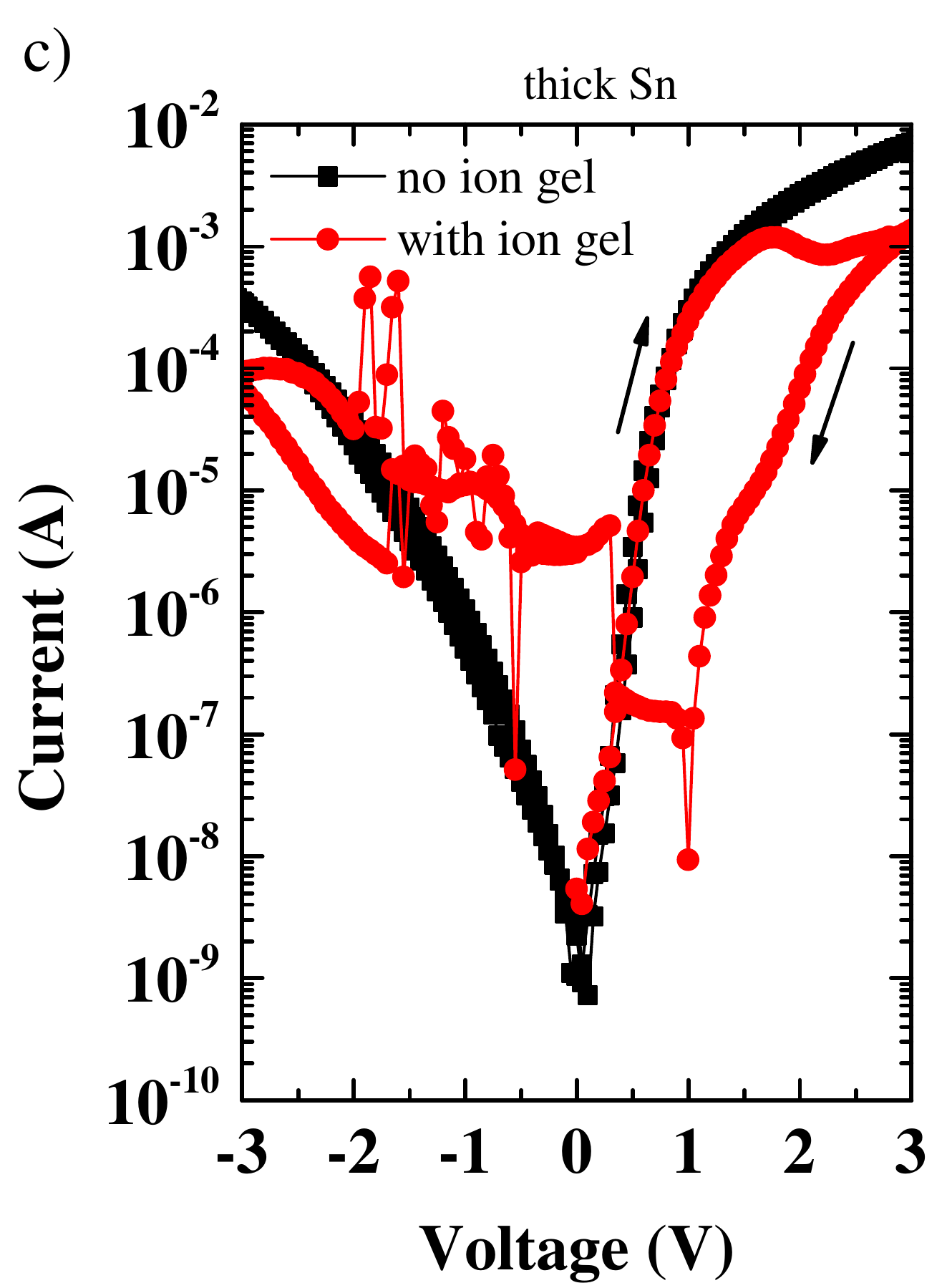}
\includegraphics[width=0.47\columnwidth]{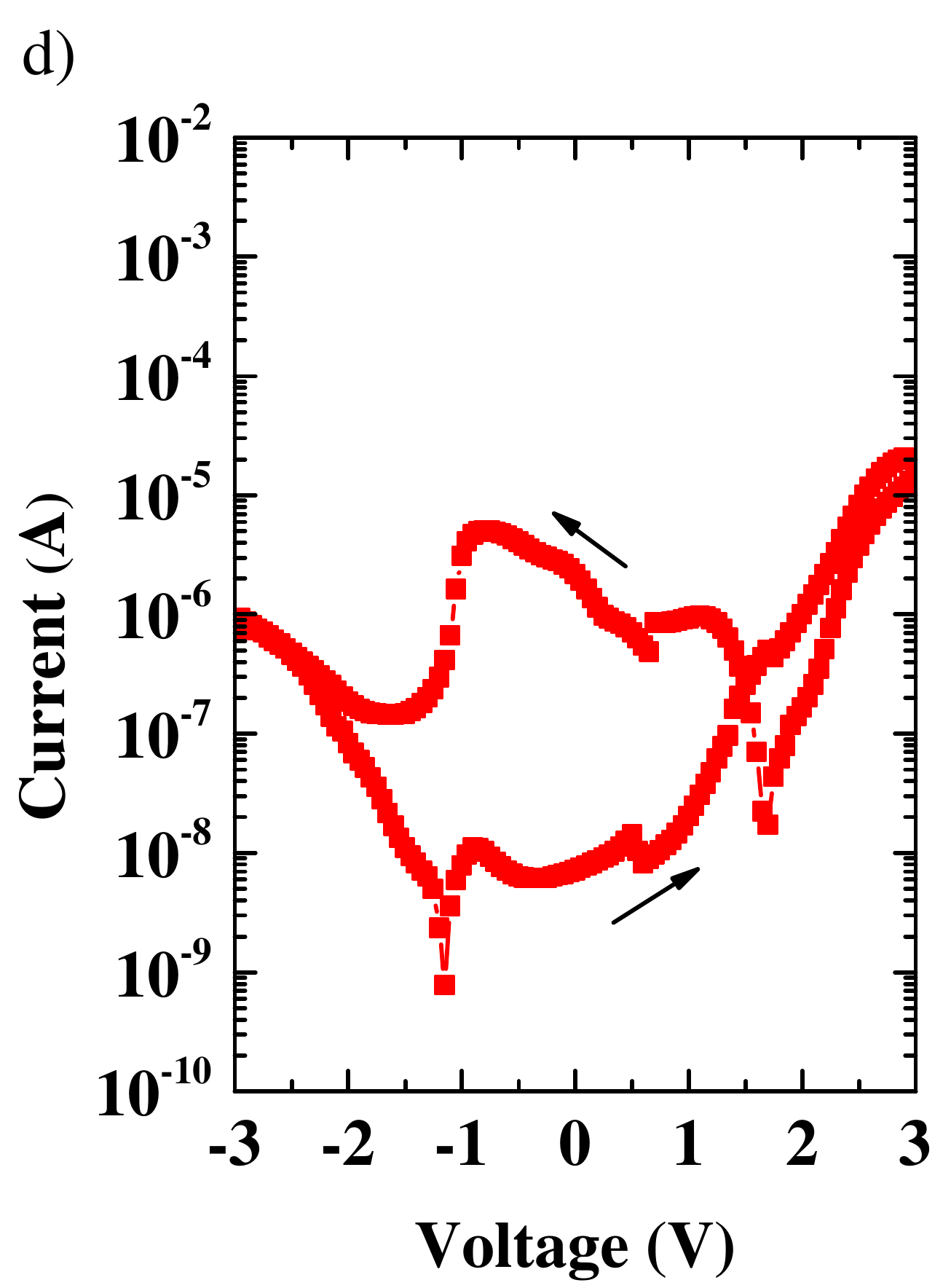}
\end{center}
\caption{\label{IxV_curve} $I - V$ curves under bipolar bias with and without ion gel outer-layer for the structures, setting Sn (or ion gel) as the common electrode: (a) ITO/PBT/thin Sn; (b) ITO/PBT/intermediate Sn; (c) ITO/PBT/thick Sn; and (d) ITO/PBT/ion gel outer-layer.}
\end{figure}

 Evaporated Sn films are frequently used to produce permeable electrodes due to its grained profile. The porosity depends on its thickness, evaporation rate and roughness of the substrate 
\cite{Serbena2006,Permeable_Sn_Ivo_2015, Seidel2018_ambipolar_gate,Neri_Sn_permeable_2018}. To analyse the effect of the ion gel in contact with a metal layer, it was produced three different Sn thicknesses layers, named: (i) thin ($\sim 128\,\rm{nm}$); (ii) intermediate ($\sim 264\,\rm{nm}$) and thick ($\sim 350\,\rm{nm}$) layer. SEM images from the three different Sn thicknesses are shown in Fig.\ref{MEV_images}:(a) thin; (b) intermediate and (c) thick Sn films on PBT (left side) and on glass substrate (right side). These images were obtained at the PBT/glass frontier and close to this region. The same Sn contact displays relevant differences between grain percolation on top of PBT layer and on glass substrate. The three Sn film thicknesses on PBT semiconductor show pores along them whose profiles are: (a) low percolation paths with high porosity; (b) high percolation paths with high porosity and (c) high percolation paths with low porosity. These percolation profiles will be correlated with the $I - V$ measurements from Figure \ref{IxV_curve} in the sequence. For the three samples there is also percolation among the Sn grains on glass substrate, as desirable to produce good electrode/contact with a small amount of pores distribution. The main goal of this work is to analyze the effects of the ion gel to the organic semiconductor/metallic electrode. It is important to emphasize that the thickness of the PBT was not varied because the sequence of results that will be presented show that the effect of the ion gel outer-layer is on the Sn/PBT interface and not on the PBT bulk.
\par Figure \ref{IxV_curve}(a-c) shows $I - V$ curves for bipolar bias of ITO/PBT/Sn device without and with the ion gel outer-layer. From PBT polymer is expected that positive charge carriers dominate charge transport on both bias. 
The ion gel layer is composed by a polymer with dielectric properties and, in spite of that, it is possible to quantify a non-negligible electric current along the ion gel film. Therefore, it was also developed a non-usual structure for electrical characterization that is ITO/PBT/ion gel as depicted in Figure \ref{structure/architecture}(c) where the ion gel layer is used as an electrode. It will provide us a reference-line output current and its $I - V$ curve is depicted on Figure \ref{IxV_curve}(d).
\par  ITO and Sn work functions are $\sim 4.7\,\rm{eV}$ \cite{ITO_Work_function} and $\sim 4.4\,\rm{eV}$ \cite{Seidel2018_ambipolar_gate}, respectively (see flat band energy diagram in supplementary material). PBT's HOMO (highest occupied molecular orbital) and LUMO  (lowest unoccupied molecular orbital) energy levels are reported as $\sim 5.3\,\rm{eV}$ and $\sim 3.3\,\rm{eV}$, respectively \cite{Kublitski2016}. It shows an almost symmetric low energy barrier level for holes injection from both electrodes with a lower barrier for ITO as anode. The $I - V$ measurements confirms it as shown in Figure \ref{IxV_curve} (b) and (c) for structure ITO/PBT/Sn (black squares symbol). The asymmetric $I - V$ curve from \ref{IxV_curve}(b) is attributed to the grained Sn profile that reduces the active area of contact.  
\par For the non-usual structure of ITO/PBT/ion gel outer-layer, the $I-V$ measurement is depicted in Fig. \ref{IxV_curve}(d) providing us a base-line output current. The current intensity in this case is limited by the leakage current throughout the ion gel film. The same ion gel film is widely used as gate dielectric layer in electrolyte transistors \cite{Lee_ion_gel_2012, Lee2009, VET_SEIDEL2020} and it could induce the reasoning that, as an insulator, any current through it would be negligible. Therefore, this base-line output current provides one of the evidences that the ion gel film acts as an active layer. The rectifying behavior is attributed to the fact that at forward bias the ITO electrode works as an infinite charge carriers reservoir that easily inject holes though a low energy barrier. However, for reverse bias, holes are poorly drifted throughout PBT film, proportionally to the amount of the accumulated holes at the ion gel/PBT interface. Note that the accumulation layer is formed on the semiconductor surface due to the ions polarized along the ion gel under external bias with a distribution similar to that observed, e.g., in field effect transistors when ion gel is used as a dielectric layer \cite{ref2_Electrolyte_Gated_Polymer_Transistors}.
\par For samples with thin Sn thickness, Figure \ref{IxV_curve}(a), the output current intensity for ITO/PBT/Sn structure is lower than for ITO/PBT/Sn/ion gel one, at forward bias. It can be explained by the conduction along the ion gel film which is higher than along the thin Sn film with low percolation paths, depicted in SEM images (Figure \ref{MEV_images}(a)). This statement come from comparison of the output current from Figure \ref{IxV_curve}(a) and (d) (red circles symbol) that present the same output current intensities. For reverse bias, the current intensity is higher for the sample with the thin Sn contact than for ion gel as contact. That behavior is expected since, even with low percolation paths, the Sn film is metallic and contributes to the charge carrier injection, which does not happen when the ion gel is used as a contact.
\par Figure \ref{IxV_curve}(b) shows the $I-V$ curve of the device constructed with intermediate Sn thickness. The output current increases $\sim 10^3$ orders of magnitude at reverse bias and there is a small change for forward bias when compared to the sample with thin Sn thickness without ion gel. The changes are attributed to the higher number of percolation paths throughout the Sn film that provides a lower sheet resistance to the contact (see MEV images in Figure \ref{MEV_images}(b)). When the ion gel is stuck on the permeable Sn film, the output current decreases for forward bias while it increases for reverse bias, compared to the same sample with and without ion gel outer-layer. To explain this phenomenon, it is necessary to note the role of the pores in the Sn film that allows the ion gel to stay in direct contact with PBT film in some regions. In these pores region the polarized ions into the ion gel will induce charge carriers on the semiconductor surface. At forward bias, electrons are induced in these pores and will not contribute to increase the output current since PBT is a p-type semiconductor. The decrease in current intensity is attributed to an effective sheet resistance that now occurs along the Sn and ion gel films.
\par A noteworthy result to be considered on applications of any vertical architecture device whose ion gel film could be placed in contact with a permeable electrode is the analysis of the current increase under reverse bias in Figure \ref{IxV_curve}(b). The intensity of the output current depends on the effective charge carrier injection regions from the permeable electrode, formed by: (i) Sn grains acting as percolated paths when a voltage is applied and (ii) the Sn pores region where charge carriers are induced on the semiconductor surface when ions are polarized under external bias. When reverse bias is applied, anions are polarized at the pores region and positive charge carriers are induced on the semiconductor surface, forming an accumulation layer. Therefore, the output current intensity is a sum of: (i) the current from the regions where there are percolated paths along the Sn grained film (current created by charge carriers injected by the metallic Sn grains) and, (ii) the drift current from the accumulation regions into the pores. This phenomenon created by the polarized ions within the pores of the electrode can be named as self-biasing effect \cite{self_biasing_2017}, once it is equivalent to an external applied bias able to improve the output current density. 
\par This kind of exploring self-biasing could provide an alternative way to improve the density of charge carriers into the device without the need to improve the voltage range. Another technique often used to increase current density in devices is the addition of an interlayer able to improve the energy level mismatch at the injection/collector interface. Interlayers improve charge carriers injection in devices, but presents limitations due to the compatibility of sequential deposition processes like solvent or temperature as required from sequential deposition \cite{review_interlayer_tandem_structure, interlayer_step_by_step_preparation_and_self_organization_2018}. It is important to note that the microscopic effects between that created by an interlayer and the present proposal with an outer-layer are quite different. Nevertheless, an outer-layer can bring similar macroscopic results with an easy step deposition of an external dry layer.
\par
\begin{figure}[h!]
 \begin{subfigure}{0.65\columnwidth}
  \caption{} 
    \includegraphics[width=\columnwidth]{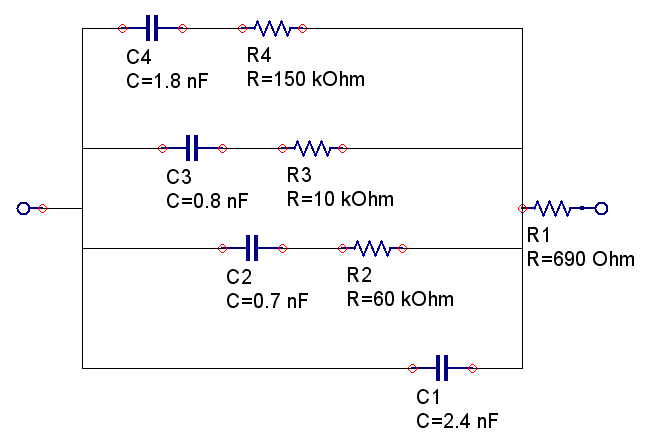}
    \end{subfigure}
    \vspace{0.5cm}
    \begin{subfigure}{0.76\columnwidth}
         \caption{} 
    \includegraphics[width=\columnwidth]{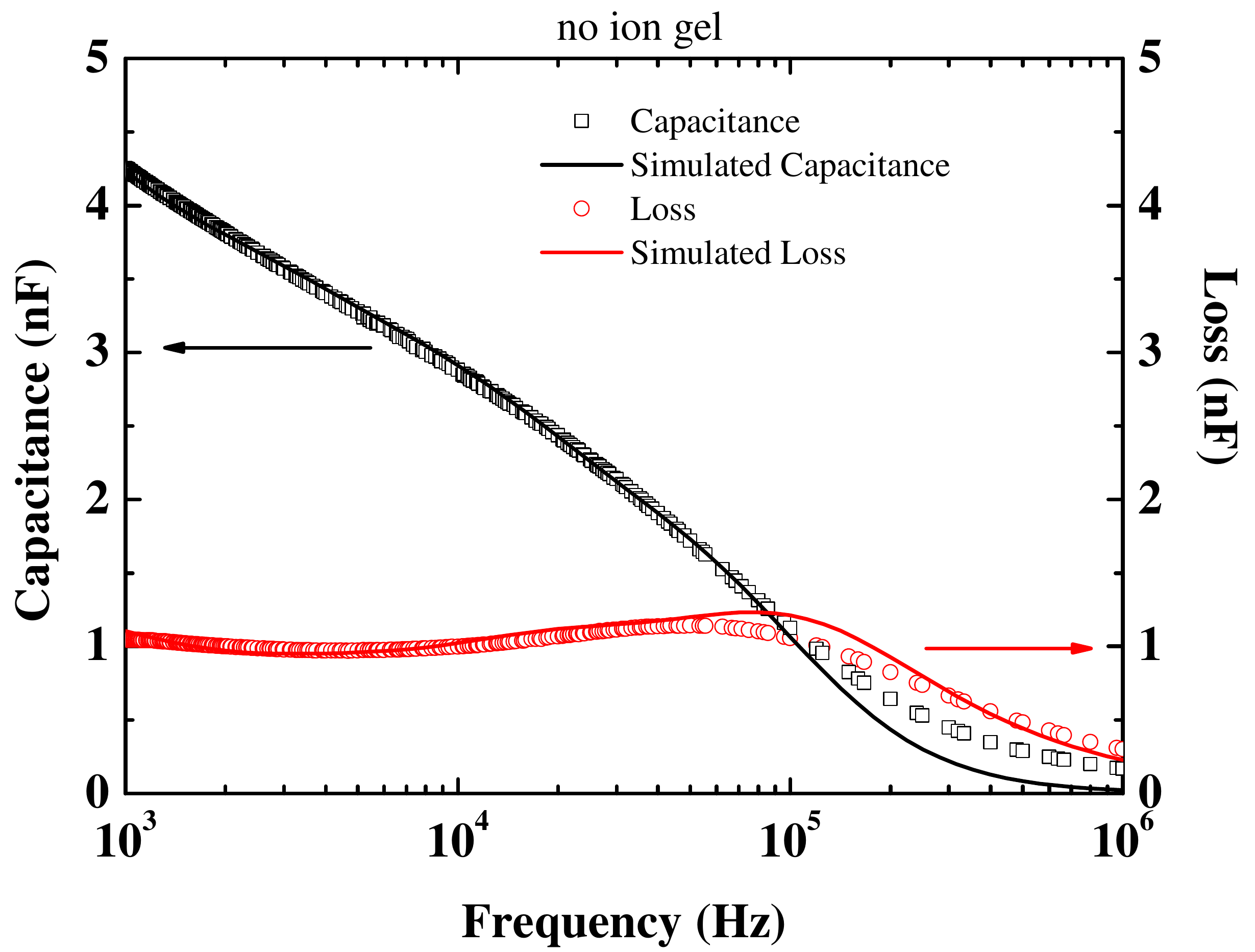}
    \end{subfigure}
      \vspace{0.5cm}
     \begin{subfigure}{0.45\columnwidth}
       \caption{}
    \includegraphics[width=\columnwidth]{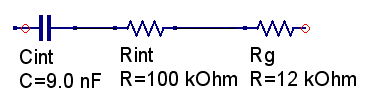}
      \end{subfigure}
        \vspace{0.5cm}
       \begin{subfigure}{0.79\columnwidth}
        \caption{}
    \includegraphics[width=\columnwidth]{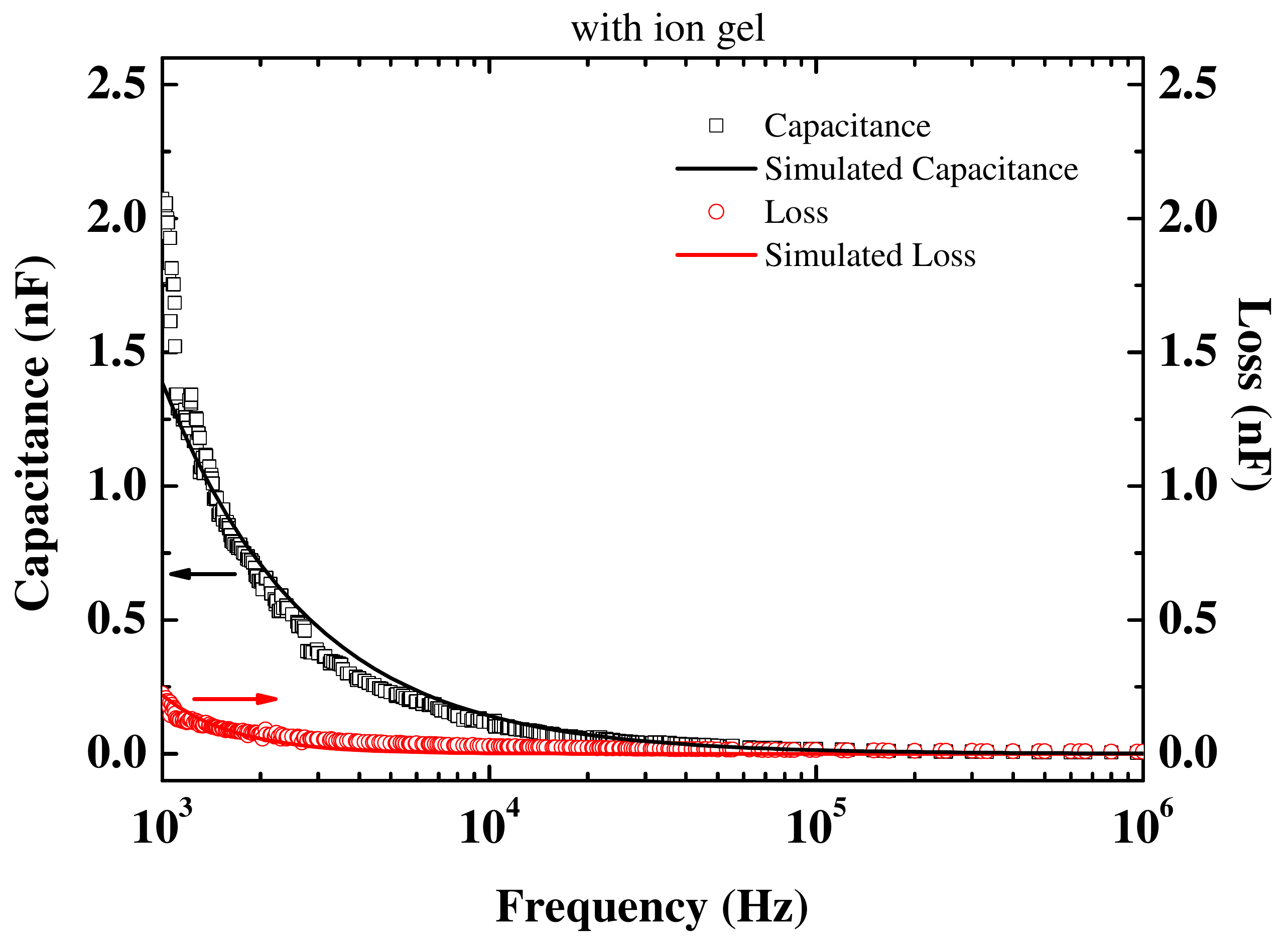}
    \end{subfigure}
      \vspace{0.5cm}
    \caption{a) Equivalent circuit and b) AC measurements and simulation of ITO/PBT/thick Sn device. c) Equivalent circuit and d) AC measurements and simulation of ITO/PBT/thick Sn/ion gel device. AC measurements are frequency dependence of the Loss and frequency dependence of the capacitance.}
    \label{simulation_fig}
\end{figure}

 Figure \ref{IxV_curve}(c) presents $I-V$ curves of the device constructed with thick Sn thickness. The output current presents similar behavior at forward bias when compared to the sample with an intermediate Sn layer, indicating that both Sn layers are good collector contacts. For reverse bias, the current is increased when compared to the sample with intermediate Sn thickness, possibly due to the higher number of percolation paths in the Sn grained film that improve the injection active area of the device under this electrical setup. When the ion gel is stuck on the sample with Sn thick film, the output current decreases for voltage ranges greater than $|2|\,\rm{V}$. The SEM images provide information of a permeable Sn film on PBT semiconductor and profilometer measurements provide the information of a Sn thickness film of $\sim 350\,\rm{nm}$. This couple of particularities bring forth the guidance of a low probability of the ion gel being in direct contact with the semiconductor into the porous. The electric field created by the polarized ions into the pores will be blocked by the thick metallic layer. However, the ionic gel layer still brings changes to the intensity of the output current, whose phenomenon that should be considered whenever an ion gel is used in contact with an electrode of any device structure. For a better understanding of the phenomenon, admittance spectroscopy was performed on these devices and equivalent circuits are proposed comparing the device with and without ion gel outer-layer. The results are showed in Figure \ref{simulation_fig}.
\par Even when very thick Sn layers was used, some electrical properties of the sandwich structure showed changes due to the ion gel outer-layer. That phenomenon are not necessarily expected once the metal contact can shield the electric field created by ions. To understand the role of the ion gel outer-layer on sandwich structure, AC measurements were performed to observe the dependence of the admittance on the frequency for the device structure with thick Sn layer and without the ion gel outer-layer. Considering these results, it is proposed an equivalent circuit simulated to extract information based on physical concepts.
\par Simulation parameters were based on experimental data provided by Souza et al., where thermally stimulated current method (TSC) was used to analyse trap levels inside PBT active layer devices\cite{Souza2014}. In their study, the authors obtained at least three different traps energy levels that support the proposed simulation parameters. Bulk and/or interface characteristics were investigated based on admittance spectroscopy method. In parallel capacitance ($C$) and Resistance ($R$), admittance is defined by \cite{stallinga2009electrical}:
\begin{equation}
 Y=G+\rm{i}\omega C\,;
 \label{eq_admitance}
\end{equation}
where $G=1/R$ is conductance, $\omega=2\pi f$ is angular frequency and $f$ is frequency. Roberts and Crowell\cite{capacitance_energy_level_1970} have described that each trap level in inorganic semiconductors acts as capacitance and resistance associated in series and any additional trap levels are associated in parallel. This information was considered in our simulations. The PBT based device without ion gel would, then, be represented as in Figure \ref{simulation_fig}(a), where $C_1$ and $R_1$ represent bulk parameters of PBT film and $C_2$, $C_3$, $C_4$ and $R_2$, $R_3$ and $R_4$ are capacitances and resistances, respectively, of the three traps energy levels provided on literature \cite{Souza2014}. The circuitry parameters are used to simulate frequency dependence of the Loss ($G/\omega$) and frequency dependence of the capacitance ($C$) curves, showed in Figure \ref{simulation_fig}(b). The simulated curve depicts good agreement with the experimental results, being the broader loss peak associated with more than one energy level for traps. It is important to note that, according to Souza et al \cite{Souza2014}, some trap levels on PBT do not have a discrete energy level as simulated in the proposed equivalent circuit of Figure \ref{simulation_fig}(a), but present a broad and possibly not homogeneously energy distribution, that can explain the small deviation between simulated and experimental curves. Since the main goal is to understand the effect of the ion gel outer-layer on the device, a more complex circuit was not simulated.
\par The admittance spectroscopy measurements for the device without and with ion gel outer-layer presented considerable changes (see, Figure\ref{simulation_fig}(b) and (d), respectively). The proposed equivalent circuit to fit the data from the sandwich structure with ion gel is shown on Figure \ref{simulation_fig}(c), where $\rm{C_{int}}$ and $\rm{R_{int}}$ are capacitance and resistance, respectively, of the PBT/Sn interface and $\rm{R_{g}}$ is associated to the effects of the ion gel outer-layer on overall device resistance. 
Experimental and simulated data have good agreement. Simulated parameters provide the information that now the interface dominates the electrical properties of the device in contrast to the previous one. From these results only a possible relaxation was fitted at low frequencies due to the dominant effect of interface states, changing the behavior of both capacitance and loss. It is a noteworthy result considering the thickness of Sn layer ($\sim \, 350\, \rm{nm}$) and even so the ion-gel have the capacity to dominate the resistance and suppress the effect of traps, which is a PBT bulk property, changing completely the AC electrical characteristics. Equally important is the fact that this is accomplished with no ion doping of the semiconductor, e.g. by ion diffusion from the ion gel outer-layer into PBT, as indicated by sequentially electrical measurements performed on attached and detached gel layer (see supplementary information), avoiding oxidation and degradation of the device generated by ionic species when they are used inside the sandwich structure.
\par It is plausible to suppose that these results can be extended to situations where an ion gel film is used also as a substrate (bottom outer-layer).
These results can also be considered on many other devices such as diodes cells, solar cells and even as a partial structure of vertical electrolyte transistors since the architecture studied here depicts similar base geometry structure for these devices. This study aims to provide an understanding about the advantages that can be obtained with the use of an ion gel outer-layer. 
\section{Conclusions}
\par In this work it was shown that the use of an ion gel outer-layer easily stuck on a sandwich device architecture modifies its charge carriers injection in different ways depending on the thickness and porosity of the electrode that is in touch with the electrolyte film. Under the limit of Sn electrodes thicknesses analysed in this work it was observed changes in the output current intensity which means that the ion gel works as an active external layer. One result to be highlighted is the improvement in the output current for reverse bias from the sample with an intermediate Sn film promoted simply by the presence of the ion gel outer-layer. Impedance spectroscopy measurements indicate that the effects of the ion gel on the PBT/Sn interface are able to dominate the electrical characteristics of the device face to the bulk one, even for metallic electrodes as thick as $\sim 350$ nm. This easy and cheap technique can be applied in a huge variety of sandwich structure devices as a simple method to modify charge carriers injection since the ion gel can be used on top (or under a bottom) of any permeable electrode.
\section*{Supplementary Material}
\par In the supplementary material file you can find: (i) $I - V$ curves from the device before, during and after the ion gel outer-layer have been stuck on the device depicting that no ions diffuse into the semiconductor; (ii) Flat energy band diagram from the device structure of ITO/PBT/Sn; (iii) all the equations used in the equivalent circuits simulation. 
\begin{acknowledgements}
The authors thank CME/UFPR for technical support on SEM experiments. This study was financed in part by the Coordenação de Aperfeiçamento de Pessoal de Nível Superior—Brasil (CAPES)—Finance Code 001.
\end{acknowledgements}
\section*{DATA AVAILABILITY} 
The data that support the findings of this study are available from the corresponding author upon reasonable request.
%
%


\bibliography{aipsamp}

\begin{thebibliography}{32}%
\makeatletter
\providecommand \@ifxundefined [1]{%
 \@ifx{#1\undefined}
}%
\providecommand \@ifnum [1]{%
 \ifnum #1\expandafter \@firstoftwo
 \else \expandafter \@secondoftwo
 \fi
}%
\providecommand \@ifx [1]{%
 \ifx #1\expandafter \@firstoftwo
 \else \expandafter \@secondoftwo
 \fi
}%
\providecommand \natexlab [1]{#1}%
\providecommand \enquote  [1]{``#1''}%
\providecommand \bibnamefont  [1]{#1}%
\providecommand \bibfnamefont [1]{#1}%
\providecommand \citenamefont [1]{#1}%
\providecommand \href@noop [0]{\@secondoftwo}%
\providecommand \href [0]{\begingroup \@sanitize@url \@href}%
\providecommand \@href[1]{\@@startlink{#1}\@@href}%
\providecommand \@@href[1]{\endgroup#1\@@endlink}%
\providecommand \@sanitize@url [0]{\catcode `\\12\catcode `\$12\catcode
  `\&12\catcode `\#12\catcode `\^12\catcode `\_12\catcode `\%12\relax}%
\providecommand \@@startlink[1]{}%
\providecommand \@@endlink[0]{}%
\providecommand \url  [0]{\begingroup\@sanitize@url \@url }%
\providecommand \@url [1]{\endgroup\@href {#1}{\urlprefix }}%
\providecommand \urlprefix  [0]{URL }%
\providecommand \Eprint [0]{\href }%
\providecommand \doibase [0]{http://dx.doi.org/}%
\providecommand \selectlanguage [0]{\@gobble}%
\providecommand \bibinfo  [0]{\@secondoftwo}%
\providecommand \bibfield  [0]{\@secondoftwo}%
\providecommand \translation [1]{[#1]}%
\providecommand \BibitemOpen [0]{}%
\providecommand \bibitemStop [0]{}%
\providecommand \bibitemNoStop [0]{.\EOS\space}%
\providecommand \EOS [0]{\spacefactor3000\relax}%
\providecommand \BibitemShut  [1]{\csname bibitem#1\endcsname}%
\let\auto@bib@innerbib\@empty
\bibitem [{\citenamefont {Zhou}\ \emph {et~al.}(2014)\citenamefont {Zhou},
  \citenamefont {Siboni}, \citenamefont {Wang}, \citenamefont {Liao},\ and\
  \citenamefont {Aziz}}]{influence_of_charge_injection}%
  \BibitemOpen
  \bibfield  {author} {\bibinfo {author} {\bibfnamefont {D.-Y.}\ \bibnamefont
  {Zhou}}, \bibinfo {author} {\bibfnamefont {H.~Z.}\ \bibnamefont {Siboni}},
  \bibinfo {author} {\bibfnamefont {Q.}~\bibnamefont {Wang}}, \bibinfo {author}
  {\bibfnamefont {L.-S.}\ \bibnamefont {Liao}}, \ and\ \bibinfo {author}
  {\bibfnamefont {H.}~\bibnamefont {Aziz}},\ }\bibfield  {title} {\enquote
  {\bibinfo {title} {The influence of charge injection from intermediate
  connectors on the performance of tandem organic light-emitting devices},}\
  }\href {\doibase 10.1063/1.4904189} {\bibfield  {journal} {\bibinfo
  {journal} {Journal of Applied Physics}\ }\textbf {\bibinfo {volume} {116}},\
  \bibinfo {pages} {223708} (\bibinfo {year} {2014})},\ \Eprint
  {http://arxiv.org/abs/https://doi.org/10.1063/1.4904189}
  {https://doi.org/10.1063/1.4904189} \BibitemShut {NoStop}%
\bibitem [{\citenamefont {Pei}\ \emph {et~al.}(1995)\citenamefont {Pei},
  \citenamefont {Yu}, \citenamefont {Zhang}, \citenamefont {Yang},\ and\
  \citenamefont {Heeger}}]{Light_Emitting_Electrochemical_Cells_Heeger}%
  \BibitemOpen
  \bibfield  {author} {\bibinfo {author} {\bibfnamefont {Q.}~\bibnamefont
  {Pei}}, \bibinfo {author} {\bibfnamefont {G.}~\bibnamefont {Yu}}, \bibinfo
  {author} {\bibfnamefont {C.}~\bibnamefont {Zhang}}, \bibinfo {author}
  {\bibfnamefont {Y.}~\bibnamefont {Yang}}, \ and\ \bibinfo {author}
  {\bibfnamefont {A.~J.}\ \bibnamefont {Heeger}},\ }\bibfield  {title}
  {\enquote {\bibinfo {title} {Polymer light-emitting electrochemical cells},}\
  }\href {\doibase 10.1126/science.269.5227.1086} {\bibfield  {journal}
  {\bibinfo  {journal} {Science}\ }\textbf {\bibinfo {volume} {269}},\ \bibinfo
  {pages} {1086--1088} (\bibinfo {year} {1995})},\ \Eprint
  {http://arxiv.org/abs/https://science.sciencemag.org/content/269/5227/1086.full.pdf}
  {https://science.sciencemag.org/content/269/5227/1086.full.pdf} \BibitemShut
  {NoStop}%
\bibitem [{\citenamefont {Mauthner}, \citenamefont {Scherf},\ and\
  \citenamefont
  {List}(2007)}]{solid_state_electrolytes_polymer_light_emitting_Emil}%
  \BibitemOpen
  \bibfield  {author} {\bibinfo {author} {\bibfnamefont {G.}~\bibnamefont
  {Mauthner}}, \bibinfo {author} {\bibfnamefont {U.}~\bibnamefont {Scherf}}, \
  and\ \bibinfo {author} {\bibfnamefont {E.~J.~W.}\ \bibnamefont {List}},\
  }\bibfield  {title} {\enquote {\bibinfo {title} {Cryptand based solid-state
  electrolytes in polymer light-emitting devices},}\ }\href {\doibase
  10.1063/1.2773756} {\bibfield  {journal} {\bibinfo  {journal} {Applied
  Physics Letters}\ }\textbf {\bibinfo {volume} {91}},\ \bibinfo {pages}
  {133501} (\bibinfo {year} {2007})},\ \Eprint
  {http://arxiv.org/abs/https://doi.org/10.1063/1.2773756}
  {https://doi.org/10.1063/1.2773756} \BibitemShut {NoStop}%
\bibitem [{\citenamefont {Deka}\ \emph {et~al.}(2019)\citenamefont {Deka},
  \citenamefont {Saikia}, \citenamefont {Lou}, \citenamefont {Lin},
  \citenamefont {Fang}, \citenamefont {Yang},\ and\ \citenamefont
  {Kao}}]{electrochromic_device}%
  \BibitemOpen
  \bibfield  {author} {\bibinfo {author} {\bibfnamefont {J.~R.}\ \bibnamefont
  {Deka}}, \bibinfo {author} {\bibfnamefont {D.}~\bibnamefont {Saikia}},
  \bibinfo {author} {\bibfnamefont {G.-W.}\ \bibnamefont {Lou}}, \bibinfo
  {author} {\bibfnamefont {C.-H.}\ \bibnamefont {Lin}}, \bibinfo {author}
  {\bibfnamefont {J.}~\bibnamefont {Fang}}, \bibinfo {author} {\bibfnamefont
  {Y.-C.}\ \bibnamefont {Yang}}, \ and\ \bibinfo {author} {\bibfnamefont
  {H.-M.}\ \bibnamefont {Kao}},\ }\bibfield  {title} {\enquote {\bibinfo
  {title} {Design, synthesis and characterization of polysiloxane and
  polyetherdiamine based comb-shaped hybrid solid polymer electrolytes for
  applications in electrochemical devices},}\ }\href {\doibase
  https://doi.org/10.1016/j.materresbull.2018.09.003} {\bibfield  {journal}
  {\bibinfo  {journal} {Materials Research Bulletin}\ }\textbf {\bibinfo
  {volume} {109}},\ \bibinfo {pages} {72--81} (\bibinfo {year}
  {2019})}\BibitemShut {NoStop}%
\bibitem [{\citenamefont {Nayak}\ \emph {et~al.}(2019)\citenamefont {Nayak},
  \citenamefont {Mahesh}, \citenamefont {Snaith},\ and\ \citenamefont
  {Cahen}}]{review_solar_cell}%
  \BibitemOpen
  \bibfield  {author} {\bibinfo {author} {\bibfnamefont {P.~K.}\ \bibnamefont
  {Nayak}}, \bibinfo {author} {\bibfnamefont {S.}~\bibnamefont {Mahesh}},
  \bibinfo {author} {\bibfnamefont {H.~J.}\ \bibnamefont {Snaith}}, \ and\
  \bibinfo {author} {\bibfnamefont {D.}~\bibnamefont {Cahen}},\ }\bibfield
  {title} {\enquote {\bibinfo {title} {Photovoltaic solar cell technologies:
  analysing the state of the art},}\ }\href {\doibase
  10.1038/s41578-019-0097-0} {\bibfield  {journal} {\bibinfo  {journal} {Nature
  Reviews Materials}\ }\textbf {\bibinfo {volume} {4}},\ \bibinfo {pages}
  {269--285} (\bibinfo {year} {2019})}\BibitemShut {NoStop}%
\bibitem [{\citenamefont {Seidel}\ \emph {et~al.}(2018)\citenamefont {Seidel},
  \citenamefont {Rossi}, \citenamefont {Jastrombek},\ and\ \citenamefont
  {Kalinowski}}]{Seidel2018_ambipolar_gate}%
  \BibitemOpen
  \bibfield  {author} {\bibinfo {author} {\bibfnamefont {K.~F.}\ \bibnamefont
  {Seidel}}, \bibinfo {author} {\bibfnamefont {L.}~\bibnamefont {Rossi}},
  \bibinfo {author} {\bibfnamefont {D.}~\bibnamefont {Jastrombek}}, \ and\
  \bibinfo {author} {\bibfnamefont {H.~J.}\ \bibnamefont {Kalinowski}},\
  }\bibfield  {title} {\enquote {\bibinfo {title} {Vertical organic field
  effect transistor: on--off state definition related to ambipolar gate
  biasing},}\ }\href {\doibase 10.1007/s00339-018-1982-x} {\bibfield  {journal}
  {\bibinfo  {journal} {Applied Physics A}\ }\textbf {\bibinfo {volume}
  {124}},\ \bibinfo {pages} {547} (\bibinfo {year} {2018})}\BibitemShut
  {NoStop}%
\bibitem [{\citenamefont {Ben-Sasson}\ \emph {et~al.}(2009)\citenamefont
  {Ben-Sasson}, \citenamefont {Avnon}, \citenamefont {Ploshnik}, \citenamefont
  {Globerman}, \citenamefont {Shenhar}, \citenamefont {Frey},\ and\
  \citenamefont {Tessler}}]{Tessler_patterned_electrode_lito_2009}%
  \BibitemOpen
  \bibfield  {author} {\bibinfo {author} {\bibfnamefont {A.~J.}\ \bibnamefont
  {Ben-Sasson}}, \bibinfo {author} {\bibfnamefont {E.}~\bibnamefont {Avnon}},
  \bibinfo {author} {\bibfnamefont {E.}~\bibnamefont {Ploshnik}}, \bibinfo
  {author} {\bibfnamefont {O.}~\bibnamefont {Globerman}}, \bibinfo {author}
  {\bibfnamefont {R.}~\bibnamefont {Shenhar}}, \bibinfo {author} {\bibfnamefont
  {G.~L.}\ \bibnamefont {Frey}}, \ and\ \bibinfo {author} {\bibfnamefont
  {N.}~\bibnamefont {Tessler}},\ }\bibfield  {title} {\enquote {\bibinfo
  {title} {Patterned electrode vertical field effect transistor fabricated
  using block copolymer nanotemplates},}\ }\href {\doibase 10.1063/1.3266855}
  {\bibfield  {journal} {\bibinfo  {journal} {Applied Physics Letters}\
  }\textbf {\bibinfo {volume} {95}},\ \bibinfo {pages} {213301} (\bibinfo
  {year} {2009})},\ \Eprint
  {http://arxiv.org/abs/https://doi.org/10.1063/1.3266855}
  {https://doi.org/10.1063/1.3266855} \BibitemShut {NoStop}%
\bibitem [{\citenamefont {Jiang}, \citenamefont {Gholamkhass},\ and\
  \citenamefont {Servati}(2019)}]{interlayer_2019}%
  \BibitemOpen
  \bibfield  {author} {\bibinfo {author} {\bibfnamefont {Z.}~\bibnamefont
  {Jiang}}, \bibinfo {author} {\bibfnamefont {B.}~\bibnamefont {Gholamkhass}},
  \ and\ \bibinfo {author} {\bibfnamefont {P.}~\bibnamefont {Servati}},\
  }\bibfield  {title} {\enquote {\bibinfo {title} {Effects of interlayer
  properties on the performance of tandem organic solar cells with low and high
  band gap polymers},}\ }\href {\doibase 10.1557/jmr.2019.168} {\bibfield
  {journal} {\bibinfo  {journal} {Journal of Materials Research}\ }\textbf
  {\bibinfo {volume} {34}},\ \bibinfo {pages} {2407--2415} (\bibinfo {year}
  {2019})}\BibitemShut {NoStop}%
\bibitem [{\citenamefont {Schultz}\ \emph {et~al.}(2021)\citenamefont
  {Schultz}, \citenamefont {Lungwitz}, \citenamefont {Longhi}, \citenamefont
  {Barlow}, \citenamefont {Marder},\ and\ \citenamefont
  {Koch}}]{interlayer_Norbert_2020}%
  \BibitemOpen
  \bibfield  {author} {\bibinfo {author} {\bibfnamefont {T.}~\bibnamefont
  {Schultz}}, \bibinfo {author} {\bibfnamefont {D.}~\bibnamefont {Lungwitz}},
  \bibinfo {author} {\bibfnamefont {E.}~\bibnamefont {Longhi}}, \bibinfo
  {author} {\bibfnamefont {S.}~\bibnamefont {Barlow}}, \bibinfo {author}
  {\bibfnamefont {S.~R.}\ \bibnamefont {Marder}}, \ and\ \bibinfo {author}
  {\bibfnamefont {N.}~\bibnamefont {Koch}},\ }\bibfield  {title} {\enquote
  {\bibinfo {title} {The interlayer method: A universal tool for energy level
  alignment tuning at inorganic/organic semiconductor heterojunctions},}\
  }\href {\doibase https://doi.org/10.1002/adfm.202010174} {\bibfield
  {journal} {\bibinfo  {journal} {Advanced Functional Materials}\ }\textbf
  {\bibinfo {volume} {31}},\ \bibinfo {pages} {2010174} (\bibinfo {year}
  {2021})},\ \Eprint
  {http://arxiv.org/abs/https://onlinelibrary.wiley.com/doi/pdf/10.1002/adfm.202010174}
  {https://onlinelibrary.wiley.com/doi/pdf/10.1002/adfm.202010174} \BibitemShut
  {NoStop}%
\bibitem [{\citenamefont {Kang}\ \emph {et~al.}(2015)\citenamefont {Kang},
  \citenamefont {Kee}, \citenamefont {Yu}, \citenamefont {Lee}, \citenamefont
  {Kim}, \citenamefont {Kim}, \citenamefont {Kim}, \citenamefont {Kong},\ and\
  \citenamefont {Lee}}]{interlayer_Semiconductor_semiconductor}%
  \BibitemOpen
  \bibfield  {author} {\bibinfo {author} {\bibfnamefont {H.}~\bibnamefont
  {Kang}}, \bibinfo {author} {\bibfnamefont {S.}~\bibnamefont {Kee}}, \bibinfo
  {author} {\bibfnamefont {K.}~\bibnamefont {Yu}}, \bibinfo {author}
  {\bibfnamefont {J.}~\bibnamefont {Lee}}, \bibinfo {author} {\bibfnamefont
  {G.}~\bibnamefont {Kim}}, \bibinfo {author} {\bibfnamefont {J.}~\bibnamefont
  {Kim}}, \bibinfo {author} {\bibfnamefont {J.-R.}\ \bibnamefont {Kim}},
  \bibinfo {author} {\bibfnamefont {J.}~\bibnamefont {Kong}}, \ and\ \bibinfo
  {author} {\bibfnamefont {K.}~\bibnamefont {Lee}},\ }\bibfield  {title}
  {\enquote {\bibinfo {title} {Simplified tandem polymer solar cells with an
  ideal self-organized recombination layer},}\ }\href {\doibase
  https://doi.org/10.1002/adma.201404765} {\bibfield  {journal} {\bibinfo
  {journal} {Advanced Materials}\ }\textbf {\bibinfo {volume} {27}},\ \bibinfo
  {pages} {1408--1413} (\bibinfo {year} {2015})},\ \Eprint
  {http://arxiv.org/abs/https://onlinelibrary.wiley.com/doi/pdf/10.1002/adma.201404765}
  {https://onlinelibrary.wiley.com/doi/pdf/10.1002/adma.201404765} \BibitemShut
  {NoStop}%
\bibitem [{\citenamefont {Di~Carlo~Rasi}\ and\ \citenamefont
  {Janssen}(2019)}]{review_interlayer_tandem_structure}%
  \BibitemOpen
  \bibfield  {author} {\bibinfo {author} {\bibfnamefont {D.}~\bibnamefont
  {Di~Carlo~Rasi}}\ and\ \bibinfo {author} {\bibfnamefont {R.~A.~J.}\
  \bibnamefont {Janssen}},\ }\bibfield  {title} {\enquote {\bibinfo {title}
  {Advances in solution-processed multijunction organic solar cells},}\ }\href
  {\doibase https://doi.org/10.1002/adma.201806499} {\bibfield  {journal}
  {\bibinfo  {journal} {Advanced Materials}\ }\textbf {\bibinfo {volume}
  {31}},\ \bibinfo {pages} {1806499} (\bibinfo {year} {2019})},\ \Eprint
  {http://arxiv.org/abs/https://onlinelibrary.wiley.com/doi/pdf/10.1002/adma.201806499}
  {https://onlinelibrary.wiley.com/doi/pdf/10.1002/adma.201806499} \BibitemShut
  {NoStop}%
\bibitem [{\citenamefont {Qiu}\ \emph {et~al.}(2003)\citenamefont {Qiu},
  \citenamefont {Xie}, \citenamefont {Chen}, \citenamefont {Wong},\ and\
  \citenamefont {Kwok}}]{Qiu2003}%
  \BibitemOpen
  \bibfield  {author} {\bibinfo {author} {\bibfnamefont {C.}~\bibnamefont
  {Qiu}}, \bibinfo {author} {\bibfnamefont {Z.}~\bibnamefont {Xie}}, \bibinfo
  {author} {\bibfnamefont {H.}~\bibnamefont {Chen}}, \bibinfo {author}
  {\bibfnamefont {M.}~\bibnamefont {Wong}}, \ and\ \bibinfo {author}
  {\bibfnamefont {H.~S.}\ \bibnamefont {Kwok}},\ }\bibfield  {title} {\enquote
  {\bibinfo {title} {Comparative study of metal or oxide capped indium–tin
  oxide anodes for organic light-emitting diodes},}\ }\href {\doibase
  10.1063/1.1556184} {\bibfield  {journal} {\bibinfo  {journal} {Journal of
  Applied Physics}\ }\textbf {\bibinfo {volume} {93}},\ \bibinfo {pages}
  {3253--3258} (\bibinfo {year} {2003})},\ \Eprint
  {http://arxiv.org/abs/https://doi.org/10.1063/1.1556184}
  {https://doi.org/10.1063/1.1556184} \BibitemShut {NoStop}%
\bibitem [{\citenamefont {Cattin}\ \emph {et~al.}(2009)\citenamefont {Cattin},
  \citenamefont {Dahou}, \citenamefont {Lare}, \citenamefont {Morsli},
  \citenamefont {Tricot}, \citenamefont {Houari}, \citenamefont {Mokrani},
  \citenamefont {Jondo}, \citenamefont {Khelil}, \citenamefont {Napo},\ and\
  \citenamefont {Bernède}}]{Cattin2009}%
  \BibitemOpen
  \bibfield  {author} {\bibinfo {author} {\bibfnamefont {L.}~\bibnamefont
  {Cattin}}, \bibinfo {author} {\bibfnamefont {F.}~\bibnamefont {Dahou}},
  \bibinfo {author} {\bibfnamefont {Y.}~\bibnamefont {Lare}}, \bibinfo {author}
  {\bibfnamefont {M.}~\bibnamefont {Morsli}}, \bibinfo {author} {\bibfnamefont
  {R.}~\bibnamefont {Tricot}}, \bibinfo {author} {\bibfnamefont
  {S.}~\bibnamefont {Houari}}, \bibinfo {author} {\bibfnamefont
  {A.}~\bibnamefont {Mokrani}}, \bibinfo {author} {\bibfnamefont
  {K.}~\bibnamefont {Jondo}}, \bibinfo {author} {\bibfnamefont
  {A.}~\bibnamefont {Khelil}}, \bibinfo {author} {\bibfnamefont
  {K.}~\bibnamefont {Napo}}, \ and\ \bibinfo {author} {\bibfnamefont {J.~C.}\
  \bibnamefont {Bernède}},\ }\bibfield  {title} {\enquote {\bibinfo {title}
  {Moo3 surface passivation of the transparent anode in organic solar cells
  using ultrathin films},}\ }\href {\doibase 10.1063/1.3077160} {\bibfield
  {journal} {\bibinfo  {journal} {Journal of Applied Physics}\ }\textbf
  {\bibinfo {volume} {105}},\ \bibinfo {pages} {034507} (\bibinfo {year}
  {2009})},\ \Eprint {http://arxiv.org/abs/https://doi.org/10.1063/1.3077160}
  {https://doi.org/10.1063/1.3077160} \BibitemShut {NoStop}%
\bibitem [{\citenamefont {Yang}\ \emph {et~al.}(2019)\citenamefont {Yang},
  \citenamefont {Zhang}, \citenamefont {Li}, \citenamefont {Yao}, \citenamefont
  {Li},\ and\ \citenamefont
  {Hou}}]{interlayer_step_by_step_preparation_and_self_organization_2018}%
  \BibitemOpen
  \bibfield  {author} {\bibinfo {author} {\bibfnamefont {B.}~\bibnamefont
  {Yang}}, \bibinfo {author} {\bibfnamefont {S.}~\bibnamefont {Zhang}},
  \bibinfo {author} {\bibfnamefont {S.}~\bibnamefont {Li}}, \bibinfo {author}
  {\bibfnamefont {H.}~\bibnamefont {Yao}}, \bibinfo {author} {\bibfnamefont
  {W.}~\bibnamefont {Li}}, \ and\ \bibinfo {author} {\bibfnamefont
  {J.}~\bibnamefont {Hou}},\ }\bibfield  {title} {\enquote {\bibinfo {title} {A
  self-organized poly(vinylpyrrolidone)-based cathode interlayer in inverted
  fullerene-free organic solar cells},}\ }\href {\doibase
  10.1002/adma.201804657} {\bibfield  {journal} {\bibinfo  {journal} {Advanced
  Materials}\ }\textbf {\bibinfo {volume} {31}},\ \bibinfo {pages} {1804657}
  (\bibinfo {year} {2019})},\ \Eprint
  {http://arxiv.org/abs/https://onlinelibrary.wiley.com/doi/pdf/10.1002/adma.201804657}
  {https://onlinelibrary.wiley.com/doi/pdf/10.1002/adma.201804657} \BibitemShut
  {NoStop}%
\bibitem [{\citenamefont {Seidel}\ \emph {et~al.}(2020)\citenamefont {Seidel},
  \citenamefont {Lungwitz}, \citenamefont {Opitz}, \citenamefont {Krüger},
  \citenamefont {Behrends}, \citenamefont {Marder},\ and\ \citenamefont
  {Koch}}]{Seidel_2020_cathode_interlayer}%
  \BibitemOpen
  \bibfield  {author} {\bibinfo {author} {\bibfnamefont {K.~F.}\ \bibnamefont
  {Seidel}}, \bibinfo {author} {\bibfnamefont {D.}~\bibnamefont {Lungwitz}},
  \bibinfo {author} {\bibfnamefont {A.}~\bibnamefont {Opitz}}, \bibinfo
  {author} {\bibfnamefont {T.}~\bibnamefont {Krüger}}, \bibinfo {author}
  {\bibfnamefont {J.}~\bibnamefont {Behrends}}, \bibinfo {author}
  {\bibfnamefont {S.~R.}\ \bibnamefont {Marder}}, \ and\ \bibinfo {author}
  {\bibfnamefont {N.}~\bibnamefont {Koch}},\ }\bibfield  {title} {\enquote
  {\bibinfo {title} {Single-step formation of a low work function cathode
  interlayer and n-type bulk doping from semiconducting
  polymer/polyethylenimine blend solution},}\ }\href {\doibase
  10.1021/acsami.0c05857} {\bibfield  {journal} {\bibinfo  {journal} {ACS
  Applied Materials \& Interfaces}\ }\textbf {\bibinfo {volume} {12}},\
  \bibinfo {pages} {28801--28807} (\bibinfo {year} {2020})},\ \bibinfo {note}
  {pMID: 32462863},\ \Eprint
  {http://arxiv.org/abs/https://doi.org/10.1021/acsami.0c05857}
  {https://doi.org/10.1021/acsami.0c05857} \BibitemShut {NoStop}%
\bibitem [{\citenamefont {Zhang}\ \emph {et~al.}(2019)\citenamefont {Zhang},
  \citenamefont {Zhang}, \citenamefont {Zhou}, \citenamefont {Tanaka},
  \citenamefont {Fong},\ and\ \citenamefont
  {Ramanathan}}]{ionic_electronic_doping_2018}%
  \BibitemOpen
  \bibfield  {author} {\bibinfo {author} {\bibfnamefont {H.-T.}\ \bibnamefont
  {Zhang}}, \bibinfo {author} {\bibfnamefont {Z.}~\bibnamefont {Zhang}},
  \bibinfo {author} {\bibfnamefont {H.}~\bibnamefont {Zhou}}, \bibinfo {author}
  {\bibfnamefont {H.}~\bibnamefont {Tanaka}}, \bibinfo {author} {\bibfnamefont
  {D.~D.}\ \bibnamefont {Fong}}, \ and\ \bibinfo {author} {\bibfnamefont
  {S.}~\bibnamefont {Ramanathan}},\ }\bibfield  {title} {\enquote {\bibinfo
  {title} {Beyond electrostatic modification: design and discovery of
  functional oxide phases via ionic-electronic doping},}\ }\href {\doibase
  10.1080/23746149.2018.1523686} {\bibfield  {journal} {\bibinfo  {journal}
  {Advances in Physics: X}\ }\textbf {\bibinfo {volume} {4}},\ \bibinfo {pages}
  {1523686} (\bibinfo {year} {2019})},\ \Eprint
  {http://arxiv.org/abs/https://doi.org/10.1080/23746149.2018.1523686}
  {https://doi.org/10.1080/23746149.2018.1523686} \BibitemShut {NoStop}%
\bibitem [{\citenamefont {Yuen}\ \emph {et~al.}(2007)\citenamefont {Yuen},
  \citenamefont {Dhoot}, \citenamefont {Namdas}, \citenamefont {Coates},
  \citenamefont {Heeney}, \citenamefont {McCulloch}, \citenamefont {Moses},\
  and\ \citenamefont {Heeger}}]{ref2_Electrolyte_Gated_Polymer_Transistors}%
  \BibitemOpen
  \bibfield  {author} {\bibinfo {author} {\bibfnamefont {J.~D.}\ \bibnamefont
  {Yuen}}, \bibinfo {author} {\bibfnamefont {A.~S.}\ \bibnamefont {Dhoot}},
  \bibinfo {author} {\bibfnamefont {E.~B.}\ \bibnamefont {Namdas}}, \bibinfo
  {author} {\bibfnamefont {N.~E.}\ \bibnamefont {Coates}}, \bibinfo {author}
  {\bibfnamefont {M.}~\bibnamefont {Heeney}}, \bibinfo {author} {\bibfnamefont
  {I.}~\bibnamefont {McCulloch}}, \bibinfo {author} {\bibfnamefont
  {D.}~\bibnamefont {Moses}}, \ and\ \bibinfo {author} {\bibfnamefont {A.~J.}\
  \bibnamefont {Heeger}},\ }\bibfield  {title} {\enquote {\bibinfo {title}
  {Electrochemical doping in electrolyte-gated polymer transistors},}\ }\href
  {\doibase 10.1021/ja0749845} {\bibfield  {journal} {\bibinfo  {journal}
  {Journal of the American Chemical Society}\ }\textbf {\bibinfo {volume}
  {129}},\ \bibinfo {pages} {14367--14371} (\bibinfo {year}
  {2007})}\BibitemShut {NoStop}%
\bibitem [{\citenamefont {Tang}\ \emph {et~al.}(2017)\citenamefont {Tang},
  \citenamefont {Sandstr{\"o}m}, \citenamefont {Lundberg}, \citenamefont
  {Lanz}, \citenamefont {Larsen}, \citenamefont {van Reenen}, \citenamefont
  {Kemerink},\ and\ \citenamefont {Edman}}]{ref2_Design_rules_for_LECs}%
  \BibitemOpen
  \bibfield  {author} {\bibinfo {author} {\bibfnamefont {S.}~\bibnamefont
  {Tang}}, \bibinfo {author} {\bibfnamefont {A.}~\bibnamefont {Sandstr{\"o}m}},
  \bibinfo {author} {\bibfnamefont {P.}~\bibnamefont {Lundberg}}, \bibinfo
  {author} {\bibfnamefont {T.}~\bibnamefont {Lanz}}, \bibinfo {author}
  {\bibfnamefont {C.}~\bibnamefont {Larsen}}, \bibinfo {author} {\bibfnamefont
  {S.}~\bibnamefont {van Reenen}}, \bibinfo {author} {\bibfnamefont
  {M.}~\bibnamefont {Kemerink}}, \ and\ \bibinfo {author} {\bibfnamefont
  {L.}~\bibnamefont {Edman}},\ }\bibfield  {title} {\enquote {\bibinfo {title}
  {Design rules for light-emitting electrochemical cells delivering bright
  luminance at 27.5 percent external quantum efficiency},}\ }\href {\doibase
  10.1038/s41467-017-01339-0} {\bibfield  {journal} {\bibinfo  {journal}
  {Nature Communications}\ }\textbf {\bibinfo {volume} {8}},\ \bibinfo {pages}
  {1190} (\bibinfo {year} {2017})}\BibitemShut {NoStop}%
\bibitem [{\citenamefont {Wang}\ \emph {et~al.}(2013)\citenamefont {Wang},
  \citenamefont {Lu}, \citenamefont {Xu}, \citenamefont {Kong}, \citenamefont
  {Cha}, \citenamefont {Zheng}, \citenamefont {Hsu}, \citenamefont {Yan},
  \citenamefont {Bradshaw}, \citenamefont {Prinz},\ and\ \citenamefont
  {Cui}}]{ref2_Electrochemical_tuning_of_vertically_aligned_MoS2}%
  \BibitemOpen
  \bibfield  {author} {\bibinfo {author} {\bibfnamefont {H.}~\bibnamefont
  {Wang}}, \bibinfo {author} {\bibfnamefont {Z.}~\bibnamefont {Lu}}, \bibinfo
  {author} {\bibfnamefont {S.}~\bibnamefont {Xu}}, \bibinfo {author}
  {\bibfnamefont {D.}~\bibnamefont {Kong}}, \bibinfo {author} {\bibfnamefont
  {J.~J.}\ \bibnamefont {Cha}}, \bibinfo {author} {\bibfnamefont
  {G.}~\bibnamefont {Zheng}}, \bibinfo {author} {\bibfnamefont {P.-C.}\
  \bibnamefont {Hsu}}, \bibinfo {author} {\bibfnamefont {K.}~\bibnamefont
  {Yan}}, \bibinfo {author} {\bibfnamefont {D.}~\bibnamefont {Bradshaw}},
  \bibinfo {author} {\bibfnamefont {F.~B.}\ \bibnamefont {Prinz}}, \ and\
  \bibinfo {author} {\bibfnamefont {Y.}~\bibnamefont {Cui}},\ }\bibfield
  {title} {\enquote {\bibinfo {title} {Electrochemical tuning of vertically
  aligned mos2 nanofilms and its application in improving hydrogen evolution
  reaction},}\ }\href {\doibase 10.1073/pnas.1316792110} {\bibfield  {journal}
  {\bibinfo  {journal} {Proceedings of the National Academy of Sciences}\
  }\textbf {\bibinfo {volume} {110}},\ \bibinfo {pages} {19701--19706}
  (\bibinfo {year} {2013})},\ \Eprint
  {http://arxiv.org/abs/https://www.pnas.org/content/110/49/19701.full.pdf}
  {https://www.pnas.org/content/110/49/19701.full.pdf} \BibitemShut {NoStop}%
\bibitem [{\citenamefont {Arbring~Sjöström}\ \emph
  {et~al.}(2018)\citenamefont {Arbring~Sjöström}, \citenamefont {Berggren},
  \citenamefont {Gabrielsson}, \citenamefont {Janson}, \citenamefont {Poxson},
  \citenamefont {Seitanidou},\ and\ \citenamefont
  {Simon}}]{review_A_Decade_of_Iontronic_Delivery_Devices}%
  \BibitemOpen
  \bibfield  {author} {\bibinfo {author} {\bibfnamefont {T.}~\bibnamefont
  {Arbring~Sjöström}}, \bibinfo {author} {\bibfnamefont {M.}~\bibnamefont
  {Berggren}}, \bibinfo {author} {\bibfnamefont {E.~O.}\ \bibnamefont
  {Gabrielsson}}, \bibinfo {author} {\bibfnamefont {P.}~\bibnamefont {Janson}},
  \bibinfo {author} {\bibfnamefont {D.~J.}\ \bibnamefont {Poxson}}, \bibinfo
  {author} {\bibfnamefont {M.}~\bibnamefont {Seitanidou}}, \ and\ \bibinfo
  {author} {\bibfnamefont {D.~T.}\ \bibnamefont {Simon}},\ }\bibfield  {title}
  {\enquote {\bibinfo {title} {A decade of iontronic delivery devices},}\
  }\href {\doibase 10.1002/admt.201700360} {\bibfield  {journal} {\bibinfo
  {journal} {Advanced Materials Technologies}\ }\textbf {\bibinfo {volume}
  {3}},\ \bibinfo {pages} {1700360} (\bibinfo {year} {2018})},\ \Eprint
  {http://arxiv.org/abs/https://onlinelibrary.wiley.com/doi/pdf/10.1002/admt.201700360}
  {https://onlinelibrary.wiley.com/doi/pdf/10.1002/admt.201700360} \BibitemShut
  {NoStop}%
\bibitem [{\citenamefont {Khan}\ \emph {et~al.}(2017)\citenamefont {Khan},
  \citenamefont {Rathi}, \citenamefont {Park}, \citenamefont {Lim},
  \citenamefont {Lee}, \citenamefont {Yun}, \citenamefont {Youn},\ and\
  \citenamefont {Kim}}]{self_biasing_2017}%
  \BibitemOpen
  \bibfield  {author} {\bibinfo {author} {\bibfnamefont {M.~A.}\ \bibnamefont
  {Khan}}, \bibinfo {author} {\bibfnamefont {S.}~\bibnamefont {Rathi}},
  \bibinfo {author} {\bibfnamefont {J.}~\bibnamefont {Park}}, \bibinfo {author}
  {\bibfnamefont {D.}~\bibnamefont {Lim}}, \bibinfo {author} {\bibfnamefont
  {Y.}~\bibnamefont {Lee}}, \bibinfo {author} {\bibfnamefont {S.~J.}\
  \bibnamefont {Yun}}, \bibinfo {author} {\bibfnamefont {D.-H.}\ \bibnamefont
  {Youn}}, \ and\ \bibinfo {author} {\bibfnamefont {G.-H.}\ \bibnamefont
  {Kim}},\ }\bibfield  {title} {\enquote {\bibinfo {title} {Junctionless diode
  enabled by self-bias effect of ion gel in single-layer mos2 device},}\ }\href
  {\doibase 10.1021/acsami.7b06071} {\bibfield  {journal} {\bibinfo  {journal}
  {ACS Applied Materials {\&} Interfaces}\ }\textbf {\bibinfo {volume} {9}},\
  \bibinfo {pages} {26983--26989} (\bibinfo {year} {2017})}\BibitemShut
  {NoStop}%
\bibitem [{\citenamefont {Souza}\ \emph {et~al.}(2014)\citenamefont {Souza},
  \citenamefont {Kowalski}, \citenamefont {Akcelrud},\ and\ \citenamefont
  {Serbena}}]{Souza2014}%
  \BibitemOpen
  \bibfield  {author} {\bibinfo {author} {\bibfnamefont {J.~d. F.~P.}\
  \bibnamefont {Souza}}, \bibinfo {author} {\bibfnamefont {E.~L.}\ \bibnamefont
  {Kowalski}}, \bibinfo {author} {\bibfnamefont {L.~C.}\ \bibnamefont
  {Akcelrud}}, \ and\ \bibinfo {author} {\bibfnamefont {J.~P.~M.}\ \bibnamefont
  {Serbena}},\ }\bibfield  {title} {\enquote {\bibinfo {title}
  {Magnetoresistance in electrochemically deposited polybithiophene thin
  films},}\ }\href {\doibase 10.1007/s10008-014-2576-y} {\bibfield  {journal}
  {\bibinfo  {journal} {Journal of Solid State Electrochemistry}\ }\textbf
  {\bibinfo {volume} {18}},\ \bibinfo {pages} {3491--3497} (\bibinfo {year}
  {2014})}\BibitemShut {NoStop}%
\bibitem [{\citenamefont {Hyung}\ \emph {et~al.}(2012)\citenamefont {Hyung},
  \citenamefont {Sung}, \citenamefont {Sipei}, \citenamefont {Yuanyan},
  \citenamefont {P.},\ and\ \citenamefont {Daniel}}]{Lee_ion_gel_2012}%
  \BibitemOpen
  \bibfield  {author} {\bibinfo {author} {\bibfnamefont {L.~K.}\ \bibnamefont
  {Hyung}}, \bibinfo {author} {\bibfnamefont {K.~M.}\ \bibnamefont {Sung}},
  \bibinfo {author} {\bibfnamefont {Z.}~\bibnamefont {Sipei}}, \bibinfo
  {author} {\bibfnamefont {G.}~\bibnamefont {Yuanyan}}, \bibinfo {author}
  {\bibfnamefont {L.~T.}\ \bibnamefont {P.}}, \ and\ \bibinfo {author}
  {\bibfnamefont {F.~C.}\ \bibnamefont {Daniel}},\ }\bibfield  {title}
  {\enquote {\bibinfo {title} {“cut and stick” rubbery ion gels as high
  capacitance gate dielectrics},}\ }\href {\doibase 10.1002/adma.201200950}
  {\bibfield  {journal} {\bibinfo  {journal} {Advanced Materials}\ }\textbf
  {\bibinfo {volume} {24}},\ \bibinfo {pages} {4457--4462} (\bibinfo {year}
  {2012})},\ \Eprint
  {http://arxiv.org/abs/https://onlinelibrary.wiley.com/doi/pdf/10.1002/adma.201200950}
  {https://onlinelibrary.wiley.com/doi/pdf/10.1002/adma.201200950} \BibitemShut
  {NoStop}%
\bibitem [{\citenamefont {Serbena}\ \emph {et~al.}(2006)\citenamefont
  {Serbena}, \citenamefont {Hümmelgen}, \citenamefont {Hadizad},\ and\
  \citenamefont {Wang}}]{Serbena2006}%
  \BibitemOpen
  \bibfield  {author} {\bibinfo {author} {\bibfnamefont {J.}~\bibnamefont
  {Serbena}}, \bibinfo {author} {\bibfnamefont {I.}~\bibnamefont {Hümmelgen}},
  \bibinfo {author} {\bibfnamefont {T.}~\bibnamefont {Hadizad}}, \ and\
  \bibinfo {author} {\bibfnamefont {Z.}~\bibnamefont {Wang}},\ }\bibfield
  {title} {\enquote {\bibinfo {title} {Hybrid permeable-base transistors based
  on an indenofluorene derivative},}\ }\href {\doibase 10.1002/smll.200500305}
  {\bibfield  {journal} {\bibinfo  {journal} {Small}\ }\textbf {\bibinfo
  {volume} {2}},\ \bibinfo {pages} {372--374} (\bibinfo {year} {2006})},\
  \Eprint
  {http://arxiv.org/abs/https://onlinelibrary.wiley.com/doi/pdf/10.1002/smll.200500305}
  {https://onlinelibrary.wiley.com/doi/pdf/10.1002/smll.200500305} \BibitemShut
  {NoStop}%
\bibitem [{\citenamefont {Kvitschal}, \citenamefont {Cruz-Cruz},\ and\
  \citenamefont {Hummelgen}(2015)}]{Permeable_Sn_Ivo_2015}%
  \BibitemOpen
  \bibfield  {author} {\bibinfo {author} {\bibfnamefont {A.}~\bibnamefont
  {Kvitschal}}, \bibinfo {author} {\bibfnamefont {I.}~\bibnamefont
  {Cruz-Cruz}}, \ and\ \bibinfo {author} {\bibfnamefont {I.~A.}\ \bibnamefont
  {Hummelgen}},\ }\bibfield  {title} {\enquote {\bibinfo {title} {Copper
  phthalocyanine based vertical organic field effect transistor with naturally
  patterned tin intermediate grid electrode},}\ }\href {\doibase
  https://doi.org/10.1016/j.orgel.2015.09.010} {\bibfield  {journal} {\bibinfo
  {journal} {Organic Electronics}\ }\textbf {\bibinfo {volume} {27}},\ \bibinfo
  {pages} {155 -- 159} (\bibinfo {year} {2015})}\BibitemShut {NoStop}%
\bibitem [{\citenamefont {Nogueira}\ \emph {et~al.}(2018)\citenamefont
  {Nogueira}, \citenamefont {da~Silva~Oz{\'o}rio}, \citenamefont {da~Silva},
  \citenamefont {Morais},\ and\ \citenamefont
  {Alves}}]{Neri_Sn_permeable_2018}%
  \BibitemOpen
  \bibfield  {author} {\bibinfo {author} {\bibfnamefont {G.~L.}\ \bibnamefont
  {Nogueira}}, \bibinfo {author} {\bibfnamefont {M.}~\bibnamefont
  {da~Silva~Oz{\'o}rio}}, \bibinfo {author} {\bibfnamefont {M.~M.}\
  \bibnamefont {da~Silva}}, \bibinfo {author} {\bibfnamefont {R.~M.}\
  \bibnamefont {Morais}}, \ and\ \bibinfo {author} {\bibfnamefont
  {N.}~\bibnamefont {Alves}},\ }\bibfield  {title} {\enquote {\bibinfo {title}
  {Middle electrode in a vertical transistor structure using an sn layer by
  thermal evaporation},}\ }\href {\doibase 10.1007/s13391-018-0034-1}
  {\bibfield  {journal} {\bibinfo  {journal} {Electronic Materials Letters}\
  }\textbf {\bibinfo {volume} {14}},\ \bibinfo {pages} {319--327} (\bibinfo
  {year} {2018})}\BibitemShut {NoStop}%
\bibitem [{\citenamefont {Kim}\ \emph {et~al.}(1998)\citenamefont {Kim},
  \citenamefont {Granström}, \citenamefont {Friend}, \citenamefont
  {Johansson}, \citenamefont {Salaneck}, \citenamefont {Daik}, \citenamefont
  {Feast},\ and\ \citenamefont {Cacialli}}]{ITO_Work_function}%
  \BibitemOpen
  \bibfield  {author} {\bibinfo {author} {\bibfnamefont {J.~S.}\ \bibnamefont
  {Kim}}, \bibinfo {author} {\bibfnamefont {M.}~\bibnamefont {Granström}},
  \bibinfo {author} {\bibfnamefont {R.~H.}\ \bibnamefont {Friend}}, \bibinfo
  {author} {\bibfnamefont {N.}~\bibnamefont {Johansson}}, \bibinfo {author}
  {\bibfnamefont {W.~R.}\ \bibnamefont {Salaneck}}, \bibinfo {author}
  {\bibfnamefont {R.}~\bibnamefont {Daik}}, \bibinfo {author} {\bibfnamefont
  {W.~J.}\ \bibnamefont {Feast}}, \ and\ \bibinfo {author} {\bibfnamefont
  {F.}~\bibnamefont {Cacialli}},\ }\bibfield  {title} {\enquote {\bibinfo
  {title} {Indium–tin oxide treatments for single- and double-layer polymeric
  light-emitting diodes: The relation between the anode physical, chemical, and
  morphological properties and the device performance},}\ }\href {\doibase
  10.1063/1.368981} {\bibfield  {journal} {\bibinfo  {journal} {Journal of
  Applied Physics}\ }\textbf {\bibinfo {volume} {84}},\ \bibinfo {pages}
  {6859--6870} (\bibinfo {year} {1998})},\ \Eprint
  {http://arxiv.org/abs/https://doi.org/10.1063/1.368981}
  {https://doi.org/10.1063/1.368981} \BibitemShut {NoStop}%
\bibitem [{\citenamefont {Kublitski}\ \emph {et~al.}(2016)\citenamefont
  {Kublitski}, \citenamefont {Tavares}, \citenamefont {Serbena}, \citenamefont
  {Liu}, \citenamefont {Hu},\ and\ \citenamefont
  {H{\"u}mmelgen}}]{Kublitski2016}%
  \BibitemOpen
  \bibfield  {author} {\bibinfo {author} {\bibfnamefont {J.}~\bibnamefont
  {Kublitski}}, \bibinfo {author} {\bibfnamefont {A.~C.~B.}\ \bibnamefont
  {Tavares}}, \bibinfo {author} {\bibfnamefont {J.~P.~M.}\ \bibnamefont
  {Serbena}}, \bibinfo {author} {\bibfnamefont {Y.}~\bibnamefont {Liu}},
  \bibinfo {author} {\bibfnamefont {B.}~\bibnamefont {Hu}}, \ and\ \bibinfo
  {author} {\bibfnamefont {I.~A.}\ \bibnamefont {H{\"u}mmelgen}},\ }\bibfield
  {title} {\enquote {\bibinfo {title} {Electrode material dependent p- or
  n-like thermoelectric behavior of single electrochemically synthesized
  poly(2,2-bithiophene) layer-application to thin film thermoelectric
  generator},}\ }\href {\doibase 10.1007/s10008-016-3223-6} {\bibfield
  {journal} {\bibinfo  {journal} {Journal of Solid State Electrochemistry}\
  }\textbf {\bibinfo {volume} {20}},\ \bibinfo {pages} {2191--2196} (\bibinfo
  {year} {2016})}\BibitemShut {NoStop}%
\bibitem [{\citenamefont {Lee}\ \emph {et~al.}(2009)\citenamefont {Lee},
  \citenamefont {Kaake}, \citenamefont {Cho}, \citenamefont {Zhu},
  \citenamefont {Lodge},\ and\ \citenamefont {Frisbie}}]{Lee2009}%
  \BibitemOpen
  \bibfield  {author} {\bibinfo {author} {\bibfnamefont {J.}~\bibnamefont
  {Lee}}, \bibinfo {author} {\bibfnamefont {L.~G.}\ \bibnamefont {Kaake}},
  \bibinfo {author} {\bibfnamefont {H.~J.}\ \bibnamefont {Cho}}, \bibinfo
  {author} {\bibfnamefont {X.~Y.}\ \bibnamefont {Zhu}}, \bibinfo {author}
  {\bibfnamefont {T.~P.}\ \bibnamefont {Lodge}}, \ and\ \bibinfo {author}
  {\bibfnamefont {C.~D.}\ \bibnamefont {Frisbie}},\ }\bibfield  {title}
  {\enquote {\bibinfo {title} {{Ion gel-gated polymer thin-film transistors:
  Operating mechanism and characterization of gate dielectric capacitance,
  switching speed, and stability}},}\ }\href {\doibase 10.1021/jp901426e}
  {\bibfield  {journal} {\bibinfo  {journal} {Journal of Physical Chemistry C}\
  }\textbf {\bibinfo {volume} {113}},\ \bibinfo {pages} {8972--8981} (\bibinfo
  {year} {2009})}\BibitemShut {NoStop}%
\bibitem [{\citenamefont {Seidel}(2020)}]{VET_SEIDEL2020}%
  \BibitemOpen
  \bibfield  {author} {\bibinfo {author} {\bibfnamefont {K.~F.}\ \bibnamefont
  {Seidel}},\ }\bibfield  {title} {\enquote {\bibinfo {title} {Fabrication and
  electrical characterization of vertical electrolyte transistor},}\ }\href
  {\doibase https://doi.org/10.1016/j.cap.2020.07.012} {\bibfield  {journal}
  {\bibinfo  {journal} {Current Applied Physics}\ }\textbf {\bibinfo {volume}
  {20}},\ \bibinfo {pages} {1288 -- 1294} (\bibinfo {year} {2020})}\BibitemShut
  {NoStop}%
\bibitem [{\citenamefont {Stallinga}(2009)}]{stallinga2009electrical}%
  \BibitemOpen
  \bibfield  {author} {\bibinfo {author} {\bibfnamefont {P.}~\bibnamefont
  {Stallinga}},\ }\href@noop {} {\emph {\bibinfo {title} {Electrical
  characterization of organic electronic materials and devices}}}\ (\bibinfo
  {publisher} {Wiley Online Library},\ \bibinfo {year} {2009})\BibitemShut
  {NoStop}%
\bibitem [{\citenamefont {Roberts}\ and\ \citenamefont
  {Crowell}(1970)}]{capacitance_energy_level_1970}%
  \BibitemOpen
  \bibfield  {author} {\bibinfo {author} {\bibfnamefont {G.~I.}\ \bibnamefont
  {Roberts}}\ and\ \bibinfo {author} {\bibfnamefont {C.~R.}\ \bibnamefont
  {Crowell}},\ }\bibfield  {title} {\enquote {\bibinfo {title} {Capacitance
  energy level spectroscopy of deep‐lying semiconductor impurities using
  schottky barriers},}\ }\href {\doibase 10.1063/1.1659102} {\bibfield
  {journal} {\bibinfo  {journal} {Journal of Applied Physics}\ }\textbf
  {\bibinfo {volume} {41}},\ \bibinfo {pages} {1767--1776} (\bibinfo {year}
  {1970})},\ \Eprint {http://arxiv.org/abs/https://doi.org/10.1063/1.1659102}
  {https://doi.org/10.1063/1.1659102} \BibitemShut {NoStop}%
\end{thebibliography}%

\end{document}